\documentclass[%
 reprint,
 amsmath,amssymb,
 aps,
pra,
floatfix,
noeprint
]{revtex4-2}

\usepackage{graphicx}
\graphicspath{{figures/}} 
\usepackage{dcolumn}
\usepackage{bm}
\usepackage{hyperref}

\usepackage{subcaption}
\usepackage{cleveref}
\usepackage{color}
\renewcommand{\subref}[1]{(\ref{sub@#1})}
\newsavebox{\measurebox}

\makeatother

\begin{document}

\preprint{APS/123-QED}

\title{Ion Trapping with a Laser-written 3D Miniaturized Monolithic Linear Paul Trap for Microcavity Integration}

\author{Soon Teh}
\email{soon.teh@oist.jp}
\author{Ezra Kassa}
\author{Shaobo Gao}
\author{Shuma Oya}
\author{Hiroki Takahashi}
\email{hiroki.takahashi@oist.jp}
\affiliation{%
 Experimental Quantum Information Physics Unit \\
 Okinawa Institute of Science and Technology
}%

\date{\today}

\begin{abstract}
The miniaturization of ion trap and the precise placement of its electrodes are necessary for the integration of a microcavity to facilitate efficient ion-cavity coupling. We present a miniature monolithic ion trap made of gold-coated fused silica with high numerical aperture access. A laser writing method referred to as \textit{selective laser etching} is employed to extract a trap structure from a block of fused silica. The fully monolithic structure eliminates the need for any post-fabrication alignment. Trenches are integrated into this structure such that the various electrodes on the monolithic device remain electrically isolated following their metalization via evaporative coating. We give details of the trap design and production, along with the demonstration of successful trapping of ions and characerization of the trap.
\end{abstract}

\maketitle


\section{Introduction}
Trapped ions are among the top contenders for quantum computation and communication, satisfying the requirements outlined by DiVincenzo's Criteria \cite{DiVincenzo2000TheComputation}. Trapped ions can perform high-fidelity quantum gate operations \cite{Gaebler2016High-FidelityQubits,Srinivas2021High-fidelityQubits,Harty2014High-fidelityBit,Christensen2020High-fidelityNucleus,Clark2021High-FidelityQubits}, offer long coherence times \cite{Wang2021SingleHour}, and readily convert quantum information into ``flying qubits'' \cite{Blinov2004ObservationPhoton}. Despite these remarkable achievements, scaling beyond a few tens of quantum registers while maintaining the same control remains an ongoing pursuit. A promising approach for extending the trapped ion system is the photonic interconnect architecture \cite{Monroe2013ScalingProcessor,Monroe2014Large-scaleInterconnects}. The architecture scales by distributing entanglement between several controllable quantum modules through photonic links. Some of the major limitations associated with building the photonic architecture include the reproducibility of the module and the remote entanglement rate \cite{Bruzewicz2019Trapped-ionChallenges}.

Photon entanglement resources in a trapped ion system can be generated, for example, through the emission of polarised photons entangled with the Zeeman states of the ion \cite{Blinov2004ObservationPhoton}. A projective Bell state measurement between the flying qubits originating from two different ions then entangles the ions remotely. The remote entanglement rate between trapped ion qubit modules scales with the photon emission rate \cite{Monroe2014Large-scaleInterconnects}. To enhance the photon emission rate, one could place the ion inside an optical cavity \cite{Schupp2021InterfaceEfficiency,Krutyanskiy2019Light-matterFibre}. The enhanced photon emission probability into the cavity output scales with the cooperativity of the system, while the reduced cavity length increases the emission rate \cite{Gao2023OptimizationNetworks}. In other words, achieving a small cavity length is favorable for the remote entanglement rate \cite{EzraSimulationPaper}. Recently, strong coupling of an ion to such a microcavity with a cavity length of a few hundred micrometers was achieved \cite{Takahashi2020StrongCavity}. However, the demonstration was limited to an endcap-style trap, whose geometry is restricted to trapping only a single ion stably in the cavity field. Meanwhile, quantum information processing requires more than a few ions to achieve meaningful usage. Therefore, the next step will be to realize a trap with the ability to control multiple ions while being capable of achieving efficient ion-cavity coupling.

Ion traps capable of trapping multiple ions can be broadly separated into two categories by geometry: 3D traps (blade and wafer traps) and 2D traps (surface traps). For comparable electrode dimensions, a 3D trap produces much stronger confinement with up to one order of magnitude increment in its trap depth \cite{Podoliak2016ComparativeCavities, Xu20233D-PrintedProcessing}. The deep trapping potential will be less prone to ion loss due to background collisions. Furthermore, in our recent finding \cite{EzraSimulationPaper}, the high geometrical symmetry in a 3D trap is key for integrating a microcavity with a protective shield near the ion without adverse effects. The same cannot be said for surface traps. Thus, a 3D trap is a more suitable candidate for integrating a microcavity.

In pursuit of a scalable 3D trap, recent trends have seen proposals for mass-producible stacked wafer traps \cite{Ragg2019SegmentedWafers, Auchter2022IndustriallyDepth}. Using a semi-automatic strategy by focusing an optical microscope on the different layers and overlapping layers' images, the wafers can be aligned, leading to mismatches of only a few micrometers throughout the full wafer stack. However, due to the stacked wafer's compactness, the optical access is restricted to only the direction normal to the stack. Furthermore, the different materials that make up the different layers in the stack, as well as the adhesive, can have different thermal expansion coefficients, leading to misalignment due to uneven thermal expansion during vacuum bakeout or cryogenic cooldown. 

An alternative to using the stacked wafer traps is to consider a fully monolithic trap \cite{Wang2020CoherentlyTrap, Kiesenhofer2023ControllingTrap}, where the trap is fabricated from a single substrate. The benefit of a monolithic trap gives us the freedom of choosing a desired electrode geometry that maximizes the optical access for both the cavity and the imaging while circumventing the need for manual or assisted alignment while maintaining uniform thermal response. The main challenge for realizing such a monolithic trap lies in the electrical separation strategy \cite{Araneda2020TheOptics} to properly define the voltage of each electrode.

To our best knowledge, a miniaturized monolithic 3D trap that can efficiently couple an ion to integrate a microcavity in the transverse direction (perpendicular to the trapping axis) is yet to be demonstrated. In this article, we focus on the design, fabrication, and demonstration of such a novel monolithic 3D trap toward cavity integration. The article is organized such that we describe selective laser-induced etching (SLE) as our main fabrication technique in \ref{sec:sle}. In \ref{sec:trap} we present a novel trap design for cavity integration, then we outline the fabrication and inspection procedure for using SLE for trap fabrication. In \ref{sec:res}, we show our results in trapping an ion and characterizing the trap potential.

\section{Selective Laser-induced Etching}
\label{sec:sle}
SLE is a subtractive fabrication process in which a transparent substrate is first irradiated with femtosecond laser pulses and then submerged in an etchant that removes the volume modified by the laser. The femtosecond laser pulse is absorbed nonlinearly in the transparent medium, altering the chemical property such that the etching rate is increased compared to the unmodified volume. Due to the nonlinear effect and pulse duration shorter than the phonon coupling time ($\sim10\,$ps), the laser affected zone (LAZ) can be smaller than the actual spot size, which gives a better manufacturing resolution than continuous wave (CW) laser machining \cite{Jonusauskas20213DLaser, Sugioka2019HybridReview, He2014FemtosecondGlass}. In spite of that, to take full advantage of SLE's capability, one needs to carefully design the trap based on the practical restrictions of the SLE method. We outlined some caveats of SLE technique in the appendix \ref{app:caveats}.

On the other hand, the lithography fabrication technique is capable of producing highly precise monolithic structures on the tens of nanometer precision scale. However, some major limitations of the lithography technique are the inability to create arbitrary 3D structures and the height limitation of sub-mm. For instance, fabrication based on micro-electromechanical systems (MEMS) technology reaches a height of $100\,\mu$m \cite{Stick2005IonChip}, while the silica-on-silicon based method has succeeded in producing a 3D monolithic trap with ion-electrode separation of hundreds of micrometers \cite{Brownnutt2006MonolithicProcessors, Wilpers2012ATechnology}. Additionally, the usage of the lithography technique could lead to ion heating due to the exposed semiconductor layers \cite{Brownnutt2006MonolithicProcessors, Stick2005IonChip}, which is detrimental to quantum gate operations. 

A counterpart to the SLE technique (additive instead of subtractive fabrication) for high-precision 3D fabrication is the two-photon polymerization (2PP) \cite{Baldacchini2019Three-dimensionalPolymerization}. The 2PP process uses a femtosecond laser for a two-photon process that induces cross-linking in a photosensitive solution to form polymers in the LAZ. A functional scalable ion trap was successfully demonstrated with the 2PP technology \cite{Xu20233D-PrintedProcessing}. However, two potential drawbacks of using a polymer-based ion trap are the outgassing and the higher thermal expansion coefficient of the polymer compared to that of glass. On the other hand, printing glass with 2PP is possible with silica nanocomposite \cite{Kotz2021Two-PhotonMicrostructures} or polyhedral oligomeric silsesquioxane (POSS) resin \cite{Bauer2023AGlass}. The use of these resins requires subsequent curing at high temperatures to remove the polymer and leave behind only the fused silica. A drawback of this glass printing method is that the curing process leads to the shrinking of the final structure and may result in unwanted warping and uncontrolled dimensions. In contrast, the use of SLE in combination with our novel trap design can offer a new and reliable avenue for realizing ion traps with arbitrary geometry with high precision.

\section{The Ion Trap}
\label{sec:trap}
This section walks through our trap design bounded by the various constraints necessary for integrating a microcavity. Next, the techniques used in fabrication and the quality assessment of such a trap are discussed. And lastly, we present the integration with other components such as the oven and signal delivery method inside our vacuum chamber.

\subsection{Design}
To achieve efficient ion-cavity coupling with a trapped $\,^{40}\text{Ca}^+$ ion (our ion species of choice in this manuscript), we aim for a cavity length of $350\,\mu$m. The cavity length is compatible with a fiber Fabry-Perot cavity (FFPC) machined with $\text{CO}_2$ laser ablation with a radius of curvature on the order of a few hundred micrometers \cite{Takahashi2014NovelBirefringence}. However, trapping ions in such a microcavity can be challenging. The dielectric mirror coating on the cavity can accumulate charge over time and distort the surrounding potential \cite{KumarVerma2020ProbingIon}. Thus, we design the trap such that the stray field originating from the cavity is shielded by electrodes in front of the cavity. Constrained by the cavity length, a miniaturized trap with features at tens of micrometers with accuracy on the few micrometer scale is necessary. The geometry requirements are achievable by the aforementioned SLE technique. 

\begin{figure}[!htbp]
    \def\svgwidth{7cm}
    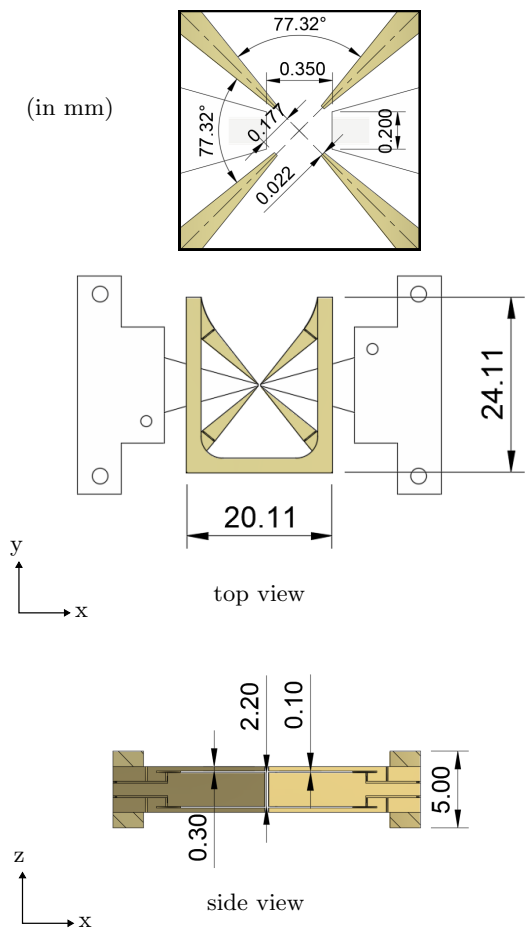
    \vspace{1mm}
    
    \caption{The geometry and dimensions of the ion trap. Top: top view of the trap with the inset showing the close-up of the trap center. The substrates on both sides of the trap have hollow channels that house the fibers with a concave mirror machined onto the facets. Bottom: side view of the trap cross-section with the substrate omitted for clarity. The given dimensions are in mm. }
    \label{fig:only_trap_dimension}
\end{figure}

\begin{figure}
    \def\svgwidth{8cm}
    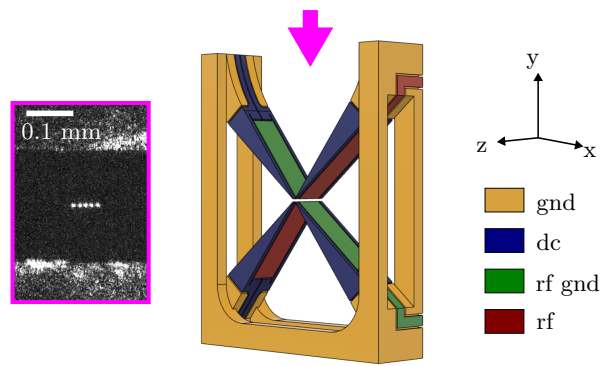
    \caption{Electric connection of the trap: A diagonal pair of the rf electrodes was driven by a signal generator at a rf frequency (\texttt{rf}) while the other pair is rf-grounded (\texttt{rf gnd}). The inset shows ions with the monolithic trap captured on a complementary metal–oxide–semiconductor (CMOS) camera, and the profile of the electrodes is also made visible by scattered radiation from a deliberately defocused laser. The magenta arrow gives the imaging direction.}
    \label{fig:simul_trap}
\end{figure}

While some of the state-of-the-art experiments have utilized hand-assembled ion traps, handling a trap with micrometer scale features is challenging and prone to misalignment. A misalignment in the trap electrode can lead to a shift in the potential and uncompensated micromotion when placed in a microcavity \cite{EzraSimulationPaper}. Restricting our design to a monolithic trap with micrometer machining tolerance circumvents the misalignment problem.

Another consideration of the design is to maximize optical access, and thus, each of the electrodes was made thin such that the numerical aperture (NA) to the ion between the radiofrequency (rf) electrodes and perpendicular to the trapping axis via $y$ axis (the magenta arrow in \cref{fig:simul_trap}) is limited only by the viewport of the vacuum chamber. The NA in our setup is currently limited to be 0.18 due to the viewport, but 0.66 when considering only the electrode geometry. To maximize axial optical access, the endcaps are constructed with four electrodes so the endcaps overlap only with the rf electrodes as shown in \cref{fig:simul_trap}.  This configuration gives an NA of 0.11 around the trap's center traversing through the ion chain along the $z$ axis. The specific trap dimensions are presented in \cref{fig:only_trap_dimension}.

In a linear ion trap, an ion is trapped by applying the dc on the endcaps (regions labeled as \texttt{dc} in \cref{fig:simul_trap} of our specific trap design) and rf voltages in the radial electrodes. In particular, we apply the rf signal on the \texttt{rf} electrodes and rf-ground the other pair (\texttt{rf gnd}) of the electrodes. Other parts of the trap (\texttt{gnd}) are grounded directly through the vacuum chamber. The different trap electrodes on our trap were electrically isolated by the presence of ``trenches'' \cite{Araneda2020TheOptics, Kiesenhofer2023ControllingTrap} which creates discontinuous conductive regions upon metalization. The exact workings of the trenches are described in the latter subsection.

\begin{figure}[htbp]
    \def\svgwidth{7cm}
    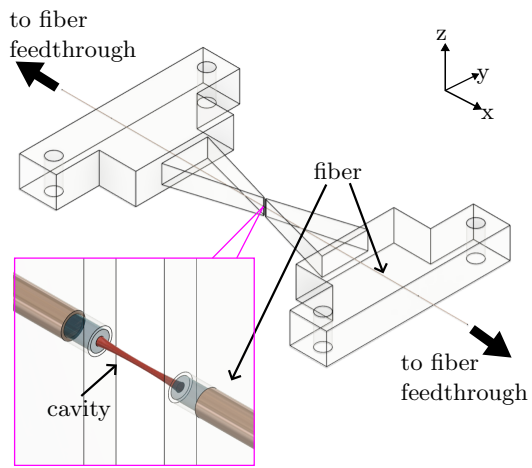
    \caption{A pair of substrates housing machined fibers is envisioned to be integrated with the trap as shown in \cref{fig:only_trap_dimension}. The fibers are inserted through hollow channels in the housing to form an optical cavity. The cavity mounting substrate should be metalized and grounded to reduce the effect of the charged fiber facets on the ion.}
    \label{fig:cavity_substrate}
\end{figure}

\begin{figure*}[!htbp]
    \centering
    \begin{subfigure}{0.3\linewidth}
    \centering
    \def\svgwidth{6cm}
\begingroup%
  \makeatletter%
  \providecommand\color[2][]{%
    \errmessage{(Inkscape) Color is used for the text in Inkscape, but the package 'color.sty' is not loaded}%
    \renewcommand\color[2][]{}%
  }%
  \providecommand\transparent[1]{%
    \errmessage{(Inkscape) Transparency is used (non-zero) for the text in Inkscape, but the package 'transparent.sty' is not loaded}%
    \renewcommand\transparent[1]{}%
  }%
  \providecommand\rotatebox[2]{#2}%
  \newcommand*\fsize{\dimexpr\f@size pt\relax}%
  \newcommand*\lineheight[1]{\fontsize{\fsize}{#1\fsize}\selectfont}%
  \ifx\svgwidth\undefined%
    \setlength{\unitlength}{731.80675597bp}%
    \ifx\svgscale\undefined%
      \relax%
    \else%
      \setlength{\unitlength}{\unitlength * \real{\svgscale}}%
    \fi%
  \else%
    \setlength{\unitlength}{\svgwidth}%
  \fi%
  \global\let\svgwidth\undefined%
  \global\let\svgscale\undefined%
  \makeatother%
  \begin{picture}(1,0.9418684)%
    \lineheight{1}%
    \setlength\tabcolsep{0pt}%
    \put(0,0){\includegraphics[width=\unitlength,page=1]{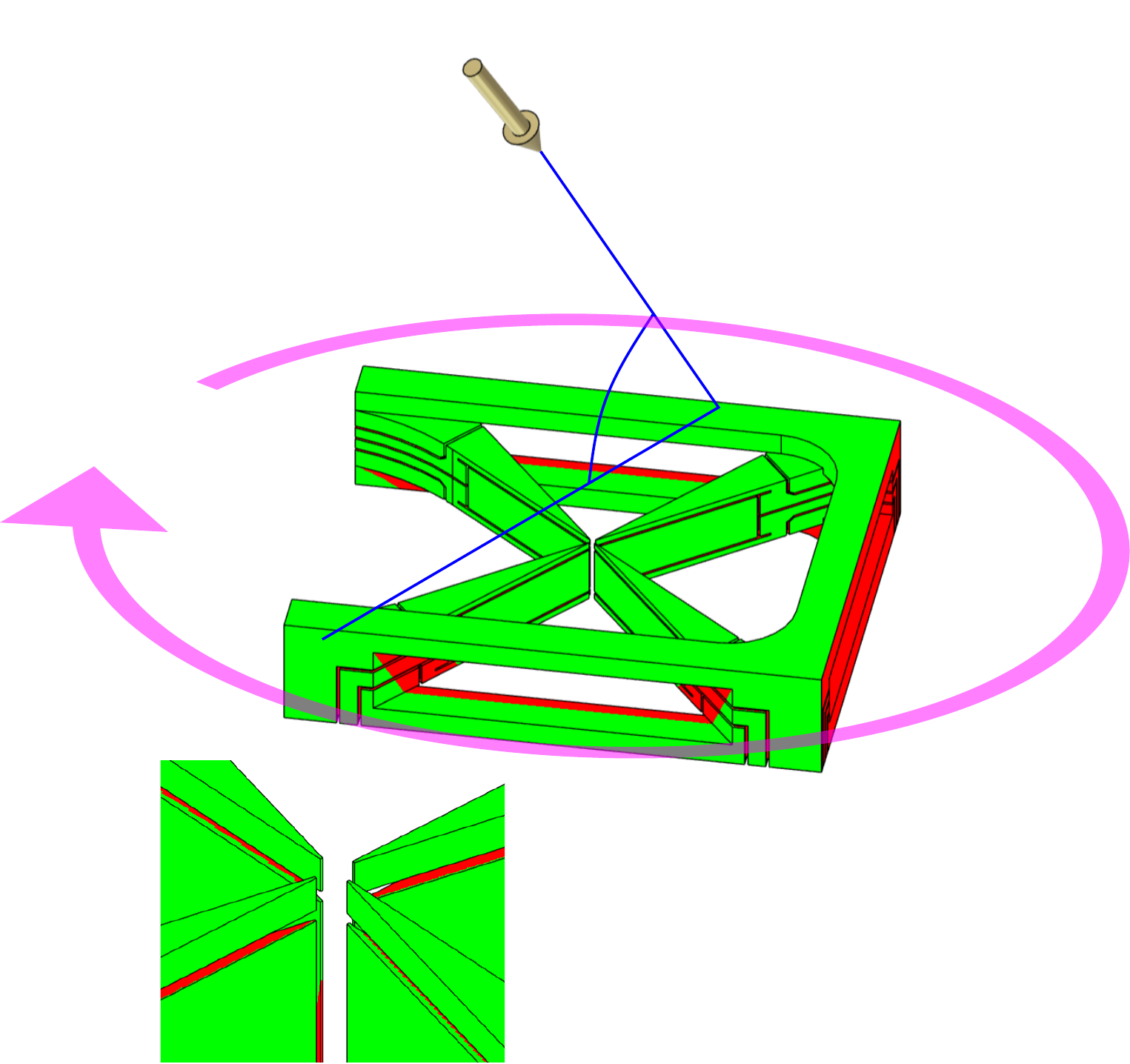}}%
    \put(0.74645627,0.23414676){\color[rgb]{1,0,1}\makebox(0,0)[lt]{\lineheight{1.25}\smash{\begin{tabular}[t]{l}rotation\end{tabular}}}}%
    \put(0,0){\includegraphics[width=\unitlength,page=2]{ebeam_angle.pdf}}%
    \put(0.63133968,0.61500855){\color[rgb]{0,0,1}\makebox(0,0)[lt]{\lineheight{1.25}\smash{\begin{tabular}[t]{l}60°\end{tabular}}}}%
    \put(0.0604927,0.89147233){\color[rgb]{0.50196078,0.50196078,0}\makebox(0,0)[lt]{\lineheight{1.25}\smash{\begin{tabular}[t]{l}evaporation\\direction\end{tabular}}}}%
  \end{picture}%
\endgroup%

    \caption{}
    \label{fig:ebeam_angle}
    \end{subfigure}
    \hfill
    \begin{subfigure}{0.6\linewidth}
    \def\svgwidth{10cm}
\begingroup%
  \makeatletter%
  \providecommand\color[2][]{%
    \errmessage{(Inkscape) Color is used for the text in Inkscape, but the package 'color.sty' is not loaded}%
    \renewcommand\color[2][]{}%
  }%
  \providecommand\transparent[1]{%
    \errmessage{(Inkscape) Transparency is used (non-zero) for the text in Inkscape, but the package 'transparent.sty' is not loaded}%
    \renewcommand\transparent[1]{}%
  }%
  \providecommand\rotatebox[2]{#2}%
  \newcommand*\fsize{\dimexpr\f@size pt\relax}%
  \newcommand*\lineheight[1]{\fontsize{\fsize}{#1\fsize}\selectfont}%
  \ifx\svgwidth\undefined%
    \setlength{\unitlength}{546.59590593bp}%
    \ifx\svgscale\undefined%
      \relax%
    \else%
      \setlength{\unitlength}{\unitlength * \real{\svgscale}}%
    \fi%
  \else%
    \setlength{\unitlength}{\svgwidth}%
  \fi%
  \global\let\svgwidth\undefined%
  \global\let\svgscale\undefined%
  \makeatother%
  \begin{picture}(1,0.55942707)%
    \lineheight{1}%
    \setlength\tabcolsep{0pt}%
    \put(0.07418903,0.39008634){\color[rgb]{0,0,0}\makebox(0,0)[lt]{\lineheight{1.25}\smash{\begin{tabular}[t]{l}e-beam evaporation \\with gold\\\end{tabular}}}}%
    \put(0.65143898,0.20939176){\color[rgb]{0,0,0}\makebox(0,0)[lt]{\lineheight{1.25}\smash{\begin{tabular}[t]{l}isolated conductive paths\end{tabular}}}}%
    \put(0,0){\includegraphics[width=\unitlength,page=1]{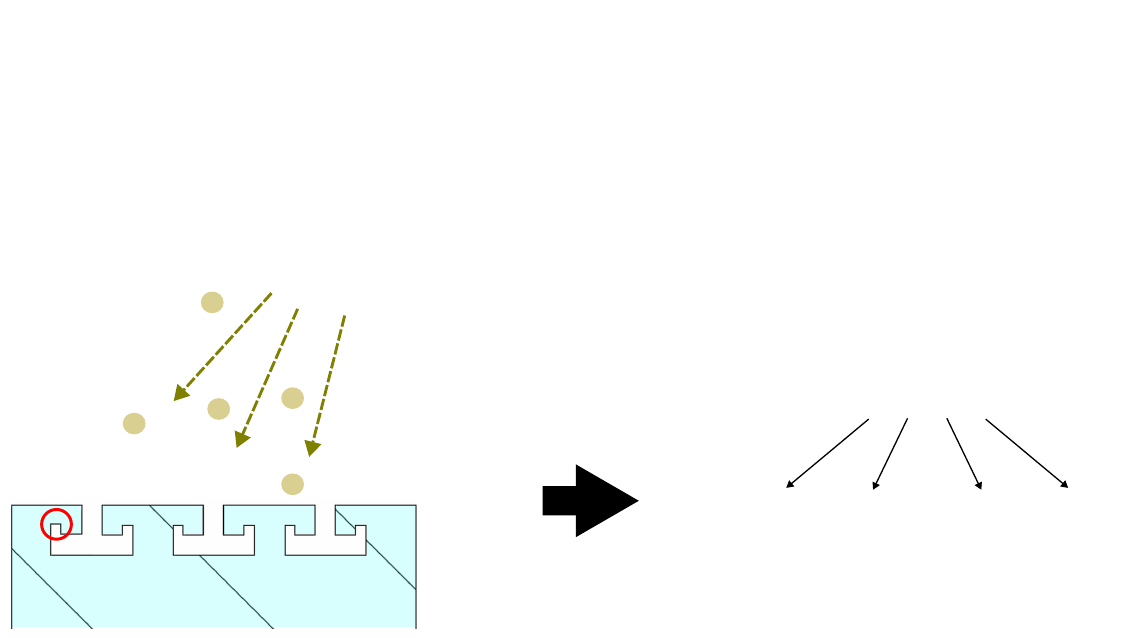}}%
    \put(-0.00117919,0.20250253){\color[rgb]{1,0,0}\makebox(0,0)[lt]{\lineheight{1.25}\smash{\begin{tabular}[t]{l}serif\end{tabular}}}}%
    \put(0,0){\includegraphics[width=\unitlength,page=2]{trap_assembly_all_drawing_ebeam_evaporation.pdf}}%
    \put(0.52469735,0.51035979){\color[rgb]{0,0,0}\makebox(0,0)[lt]{\lineheight{1.25}\smash{\begin{tabular}[t]{l}(in mm)\end{tabular}}}}%
    \put(0,0){\includegraphics[width=\unitlength,page=3]{trap_assembly_all_drawing_ebeam_evaporation.pdf}}%
  \end{picture}%
\endgroup%

    \vspace{2mm}
    \caption{}
    \label{fig:ebeam_evaporation}
    \end{subfigure}

    \caption{The evaporation process for the monolithic 3D ion trap. \subref{fig:ebeam_angle} Orientation of the trap relative to the evaporation direction. The trap is tilted and rotated throughout the evaporation to maximize the thickness of gold on the facet of the electrode tips facing the ion. The green region in the figure indicates the accessibility of the evaporation while the red regions are in the shadow of the trap. Note that they are in the shadow at a particular rotation angle. The relevant parts of the electrodes facing trapped ions are all covered with gold after a full rotation. \subref{fig:ebeam_evaporation} An example showing the principle behind using trenches to electrically isolate different electrodes. By adding serifs to the trenches, the isolation remained robust to extreme evaporation angles that are necessary for the metalization of a monolithic trap with electric connections routed in 3D. The inset shows the typical dimensions of our trench.}
    \label{fig:ebeam}
\end{figure*}

Motivated by the previous work of achieving strong ion-cavity coupling \cite{Takahashi2020StrongCavity}, we envisioned a FFPC setup as shown in \cref{fig:cavity_substrate} comprised of a pair of fiber whose facets are machined with $\text{CO}_2$ laser ablation \cite{Hunger2010AFinesse}. Each machined fiber is inserted in a substrate with a hole that fits the diameter of the fibers. For a linear trap, orienting the cavity to be perpendicular to the trapping axis could lead to large potential distortion \cite{Podoliak2016ComparativeCavities}. However, one can avoid this adverse effect by changing the driving scheme of the rf signal \cite{EzraSimulationPaper}.

A cavity's passive stability is important for maintaining a stable cavity for interfacing with the ion. The vibrational modes on the protruding tips of FFPC can lead to difficulties in locking \cite{Saavedra2021TunableStability}. To reduce the effect of vibrational noise on the cavity, the pyramidal substrate (\cref{fig:cavity_substrate}) is designed to have a large base that sits on a piezoelectric stage and a tapered tip to maximize the volume supporting the fiber at the tip. With the feature size of the cavity substrate ranging from a few hundred micrometers to a few millimeters, the SLE technique can also be employed to produce this substrate.

\subsection{Fabrication}
\begin{figure}[htbp]
    \centering
    \def\svgwidth{5cm}
    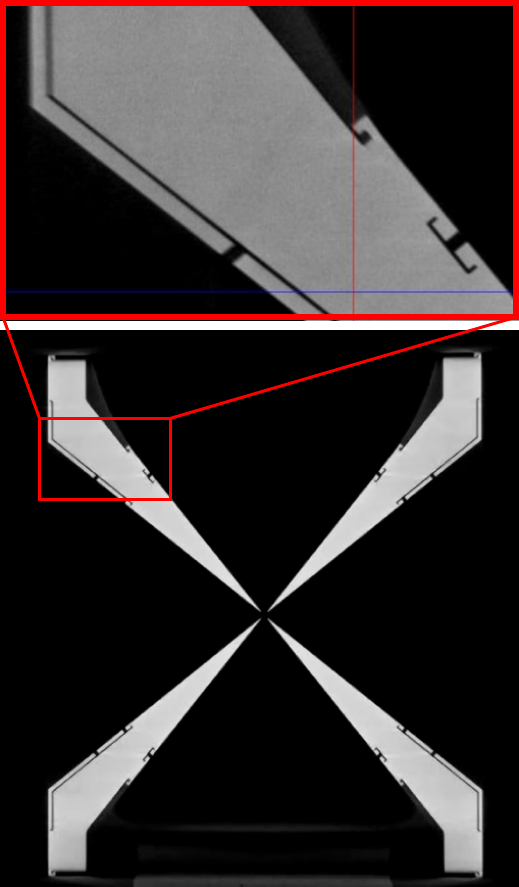
    \caption{MicroCT scans of the fabricated trap. The cross-sections of the trap can be revealed non-destructively to verify the penetration of the etching.}
    \label{fig:microct}
\end{figure}

We fabricate the trap using the SLE technique on a fused silica substrate. The fabrication details are discussed in appendix \ref{app:fab_specs}. After etching the trap, we verify the completion by checking the clearance of the trenches with micro-computed tomography (microCT) which is a nondestructive method that uses X-ray to image the cross-section of the test samples. \Cref{fig:microct} shows a slice of the reconstructed 3D image, with the inset showing good clearance for the trenches to avoid shorting the electrodes during the metalization process. The design and motivation for this trench structure will be discussed in the latter text.

The fused silica trap is metalized with e-beam evaporation ($10\,$nm of titanium as an adhesive layer followed by $2\,\mu$m of gold). The principle of separating the electrodes through the trenches is presented in \cref{fig:ebeam_evaporation}, in which the trenches create a gap between the metalized layer to define the electrodes. Due to our monolithic design, there is no single evaporation angle that can encapsulate all parts of the trap. Thus, it became necessary to investigate the evaporation angle for two reasons. First, to ensure a sufficient final gold thickness at the trap electrode facet facing the ion to prevent increased resistance due to the skin effect; Second, to ensure a closed connection for each electrical path of the electrodes. After analyzing the possibilities, we chose a 60-degree tilt and 3 rotations per minute during the evaporation process (\cref{fig:ebeam_angle}). We estimate that the effective thickness on the electrode facet is about $400\,$nm for the $2\,\mu$m thickness deposited on the base layer. This is because the evaporated beam has no continuous line of sight to all eight sides of the electrode blade edges (see \cref{fig:ebeam_angle}).  

Due to the evaporation from multiple angles, we adapted the trench design in \cite{Araneda2020TheOptics} with additional ``serifs'' as depicted in \cref{fig:ebeam_evaporation} to eliminate possibilities of short-circuits. After metalization, we tested the trap with up to $100\,$V in atmospheric pressure with no breakdown occurring. Further details on fabrication are given in appendix \ref{app:fab_specs}.

\subsection{System integration}
\begin{figure}[htbp]
    \centering
    \def\svgwidth{7cm}
    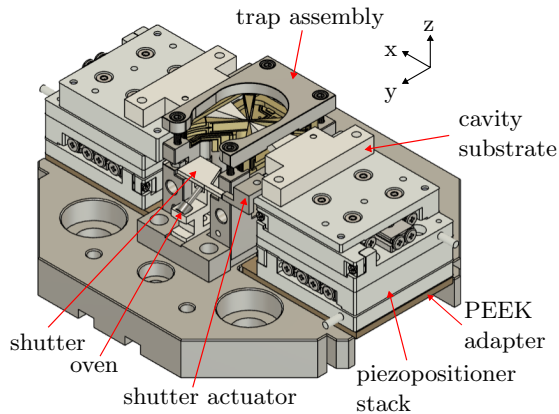
    \caption{Overview of the assembly. The trap is secured to a SLE printed intermediate circuit by a metal cap. A pair of collimated oven setups (only one of them shown) pointing towards the trap center are the calcium atom source. 
    }
    \label{fig:integration_overview}
\end{figure}

\begin{figure*}[htbp]
    \centering

    \hfill
    \sbox{\measurebox}{
    \begin{subfigure}{0.3\linewidth}
    \centering
    \vspace{5mm}
    \def\svgwidth{7cm}
\begingroup%
  \makeatletter%
  \providecommand\color[2][]{%
    \errmessage{(Inkscape) Color is used for the text in Inkscape, but the package 'color.sty' is not loaded}%
    \renewcommand\color[2][]{}%
  }%
  \providecommand\transparent[1]{%
    \errmessage{(Inkscape) Transparency is used (non-zero) for the text in Inkscape, but the package 'transparent.sty' is not loaded}%
    \renewcommand\transparent[1]{}%
  }%
  \providecommand\rotatebox[2]{#2}%
  \newcommand*\fsize{\dimexpr\f@size pt\relax}%
  \newcommand*\lineheight[1]{\fontsize{\fsize}{#1\fsize}\selectfont}%
  \ifx\svgwidth\undefined%
    \setlength{\unitlength}{247.64446589bp}%
    \ifx\svgscale\undefined%
      \relax%
    \else%
      \setlength{\unitlength}{\unitlength * \real{\svgscale}}%
    \fi%
  \else%
    \setlength{\unitlength}{\svgwidth}%
  \fi%
  \global\let\svgwidth\undefined%
  \global\let\svgscale\undefined%
  \makeatother%
  \begin{picture}(1,1.85474876)%
    \lineheight{1}%
    \setlength\tabcolsep{0pt}%
    \put(0,0){\includegraphics[width=\unitlength,page=1]{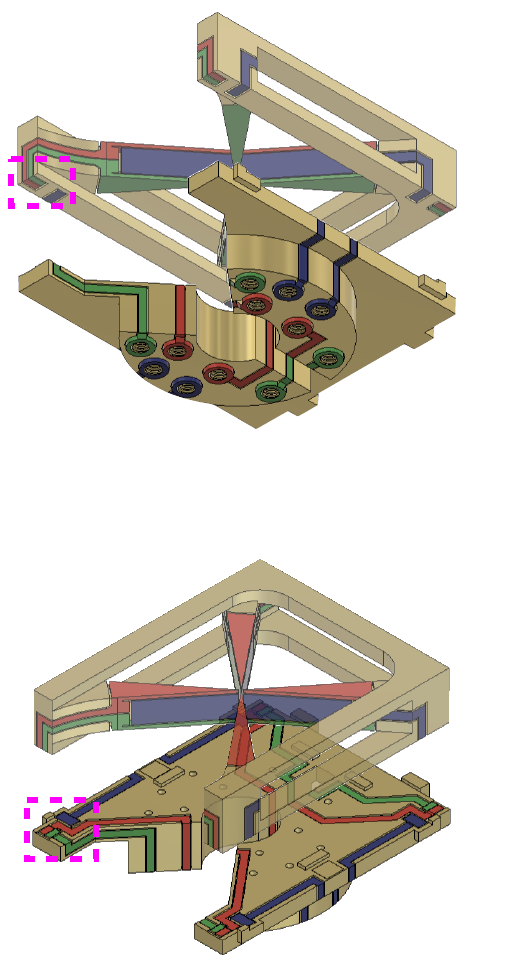}}%
    \put(-0.00324144,1.3389872){\color[rgb]{1,0,1}\makebox(0,0)[lt]{\lineheight{1.25}\smash{\begin{tabular}[t]{l}I\end{tabular}}}}%
    \put(0.01845571,0.37507833){\color[rgb]{1,0,1}\makebox(0,0)[lt]{\lineheight{1.25}\smash{\begin{tabular}[t]{l}II\end{tabular}}}}%
    \put(0,0){\includegraphics[width=\unitlength,page=2]{trap_assembly_all_drawing_assembly_multi_angle.pdf}}%
    \put(0.16262083,1.68347074){\color[rgb]{1,0,1}\makebox(0,0)[lt]{\lineheight{1.25}\smash{\begin{tabular}[t]{l}III\end{tabular}}}}%
    \put(0,0){\includegraphics[width=\unitlength,page=3]{trap_assembly_all_drawing_assembly_multi_angle.pdf}}%
  \end{picture}%
\endgroup%

    \vspace{5mm}
    \caption{}
    \label{fig:assembly_multi_angle}
    \end{subfigure}}
    \usebox{\measurebox}\qquad
    \hfill
    \begin{minipage}[b][\ht\measurebox][s]{.5\textwidth}
    \begin{subfigure}{0.8\linewidth}
    \centering
    \def\svgwidth{6cm}
\begingroup%
  \makeatletter%
  \providecommand\color[2][]{%
    \errmessage{(Inkscape) Color is used for the text in Inkscape, but the package 'color.sty' is not loaded}%
    \renewcommand\color[2][]{}%
  }%
  \providecommand\transparent[1]{%
    \errmessage{(Inkscape) Transparency is used (non-zero) for the text in Inkscape, but the package 'transparent.sty' is not loaded}%
    \renewcommand\transparent[1]{}%
  }%
  \providecommand\rotatebox[2]{#2}%
  \newcommand*\fsize{\dimexpr\f@size pt\relax}%
  \newcommand*\lineheight[1]{\fontsize{\fsize}{#1\fsize}\selectfont}%
  \ifx\svgwidth\undefined%
    \setlength{\unitlength}{392.99998847bp}%
    \ifx\svgscale\undefined%
      \relax%
    \else%
      \setlength{\unitlength}{\unitlength * \real{\svgscale}}%
    \fi%
  \else%
    \setlength{\unitlength}{\svgwidth}%
  \fi%
  \global\let\svgwidth\undefined%
  \global\let\svgscale\undefined%
  \makeatother%
  \begin{picture}(1,0.68366413)%
    \lineheight{1}%
    \setlength\tabcolsep{0pt}%
    \put(0,0){\includegraphics[width=\unitlength,page=1]{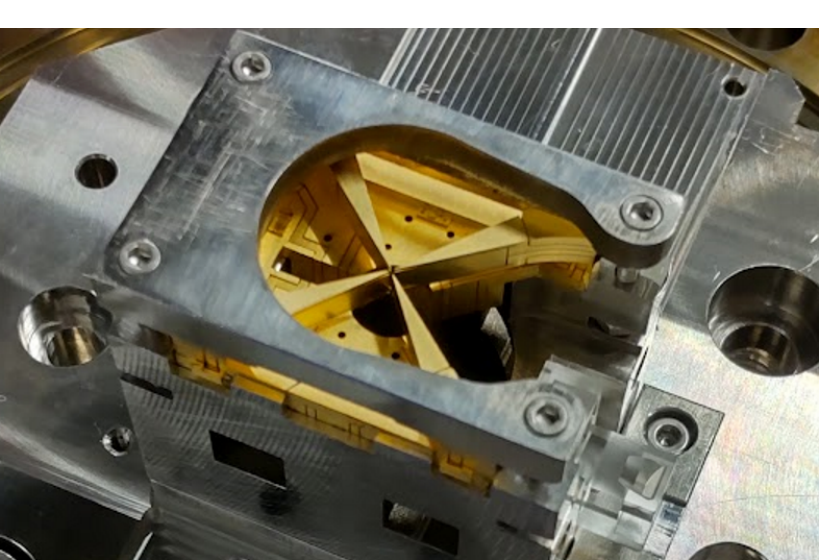}}%
    \put(0.03957118,0.11273657){\color[rgb]{0,1,0}\makebox(0,0)[lt]{\lineheight{1.25}\smash{\begin{tabular}[t]{l}fixing\\screws\end{tabular}}}}%
    \put(0,0){\includegraphics[width=\unitlength,page=2]{trap_clamped.pdf}}%
  \end{picture}%
\endgroup%

    \caption{}
    \label{fig:trap_clamped}
    \end{subfigure}
    \vfill
    \begin{subfigure}{0.8\linewidth}
    \centering
    \def\svgwidth{6cm}
\begingroup%
  \makeatletter%
  \providecommand\color[2][]{%
    \errmessage{(Inkscape) Color is used for the text in Inkscape, but the package 'color.sty' is not loaded}%
    \renewcommand\color[2][]{}%
  }%
  \providecommand\transparent[1]{%
    \errmessage{(Inkscape) Transparency is used (non-zero) for the text in Inkscape, but the package 'transparent.sty' is not loaded}%
    \renewcommand\transparent[1]{}%
  }%
  \providecommand\rotatebox[2]{#2}%
  \newcommand*\fsize{\dimexpr\f@size pt\relax}%
  \newcommand*\lineheight[1]{\fontsize{\fsize}{#1\fsize}\selectfont}%
  \ifx\svgwidth\undefined%
    \setlength{\unitlength}{417.20485898bp}%
    \ifx\svgscale\undefined%
      \relax%
    \else%
      \setlength{\unitlength}{\unitlength * \real{\svgscale}}%
    \fi%
  \else%
    \setlength{\unitlength}{\svgwidth}%
  \fi%
  \global\let\svgwidth\undefined%
  \global\let\svgscale\undefined%
  \makeatother%
  \begin{picture}(1,0.63511831)%
    \lineheight{1}%
    \setlength\tabcolsep{0pt}%
    \put(0,0){\includegraphics[width=\unitlength,page=1]{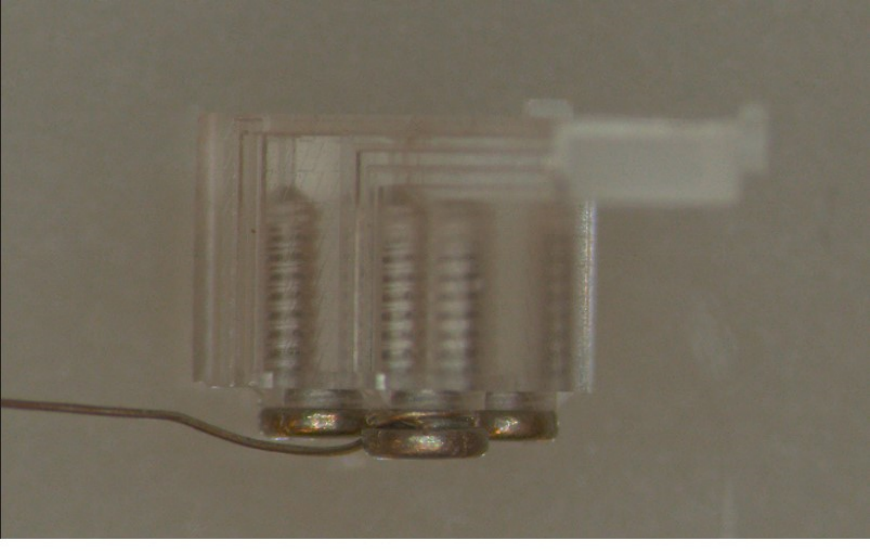}}%
    \put(0.71492035,0.05788109){\color[rgb]{0,1,0}\makebox(0,0)[lt]{\lineheight{1.25}\smash{\begin{tabular}[t]{l}screws\end{tabular}}}}%
    \put(0.06481014,0.36552451){\color[rgb]{0,1,0}\makebox(0,0)[lt]{\lineheight{1.25}\smash{\begin{tabular}[t]{l}wire\end{tabular}}}}%
    \put(0,0){\includegraphics[width=\unitlength,page=2]{wire_circuit.pdf}}%
  \end{picture}%
\endgroup%

    \caption{}
    \label{fig:wire_circuit}
    \end{subfigure}
    \vfill
    \begin{subfigure}{0.8\linewidth}
    \centering
    \vspace{3mm}
    \def\svgwidth{6cm}
\begingroup%
  \makeatletter%
  \providecommand\color[2][]{%
    \errmessage{(Inkscape) Color is used for the text in Inkscape, but the package 'color.sty' is not loaded}%
    \renewcommand\color[2][]{}%
  }%
  \providecommand\transparent[1]{%
    \errmessage{(Inkscape) Transparency is used (non-zero) for the text in Inkscape, but the package 'transparent.sty' is not loaded}%
    \renewcommand\transparent[1]{}%
  }%
  \providecommand\rotatebox[2]{#2}%
  \newcommand*\fsize{\dimexpr\f@size pt\relax}%
  \newcommand*\lineheight[1]{\fontsize{\fsize}{#1\fsize}\selectfont}%
  \ifx\svgwidth\undefined%
    \setlength{\unitlength}{568.91393947bp}%
    \ifx\svgscale\undefined%
      \relax%
    \else%
      \setlength{\unitlength}{\unitlength * \real{\svgscale}}%
    \fi%
  \else%
    \setlength{\unitlength}{\svgwidth}%
  \fi%
  \global\let\svgwidth\undefined%
  \global\let\svgscale\undefined%
  \makeatother%
  \begin{picture}(1,0.52836798)%
    \lineheight{1}%
    \setlength\tabcolsep{0pt}%
    \put(0,0){\includegraphics[width=\unitlength,page=1]{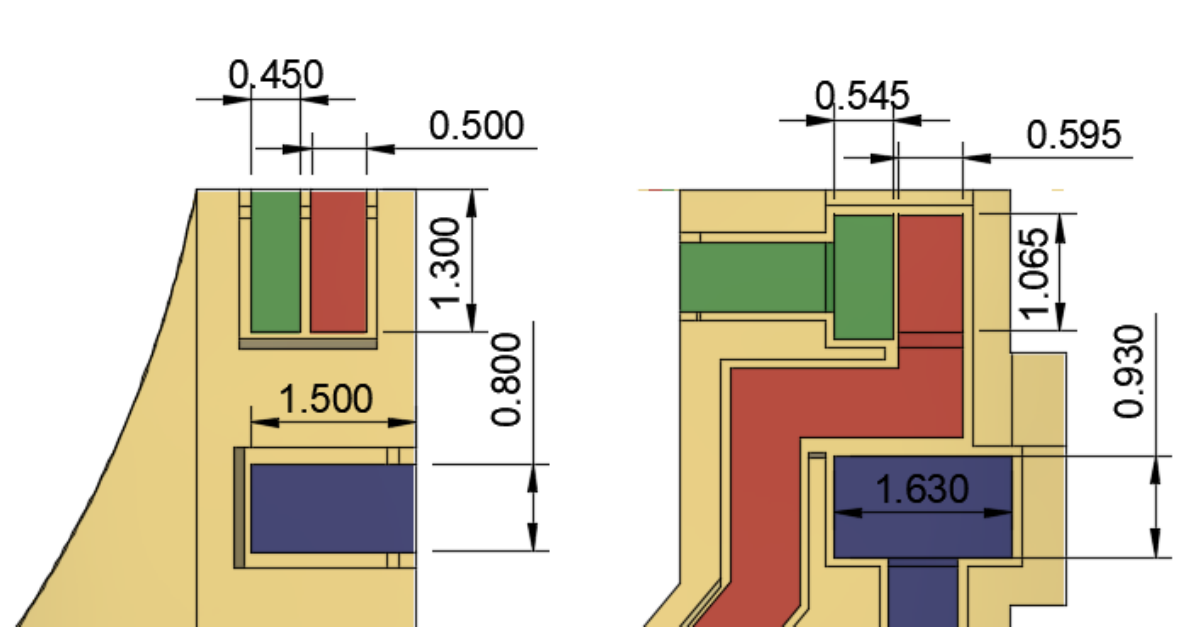}}%
    \put(0.44266685,0.50634666){\color[rgb]{0,0,0}\makebox(0,0)[lt]{\lineheight{1.25}\smash{\begin{tabular}[t]{l}(in mm)\end{tabular}}}}%
  \end{picture}%
\endgroup%

    \caption{}
    \label{fig:circuit_dimension}
    \end{subfigure}
    \end{minipage}
    \hfill

    \caption{The routing from the trap electrodes to an intermediate circuit: \subref{fig:assembly_multi_angle} shows the routing of the 3D circuit across the trap from multiple views. The false color is to guide the eye and the lowered opacity of the trap is to help distinguish the circuit component. At the bottom of the intermediate circuit, the rings are threaded holes to clamp wires with screws. The magenta boxes and arrows drawn are the regions and directions of viewing for the image and schematics in \subref{fig:wire_circuit} and \subref{fig:assembly_multi_angle} (see the following descriptions). \subref{fig:trap_clamped} shows a view of the trap assembly. The four screws at the top are used to secure the trap to the intermediate circuit. \subref{fig:wire_circuit} shows an uncoated test section of the intermediate circuit labeled III in \subref{fig:assembly_multi_angle}. The intermediate circuit is connected to the wire by a screw. The screw threads are visible through the uncoated fused silica substrate. \subref{fig:circuit_dimension} shows the dimensions of the contact pads for both the trap and the circuit, labeled as I and II in \subref{fig:assembly_multi_angle}. The figure for the trap is flipped for direct comparison. The pads on the trap side are smaller than the circuit with sufficient spacing to tolerate slight misalignment without shorting.}
    \label{fig:sle_routing}
\end{figure*}

We presented the trap design in the previous subsection, and in the following, we will delve into the details of the wider design. The overview of the integration is shown in \cref{fig:integration_overview}.

To connect the trap to a voltage source, we rely on an intermediate circuit shown in \cref{fig:assembly_multi_angle}. This intermediate circuit has electrical paths running on its surface in 3D, connecting the threaded holes at the bottom to the corresponding pads at the top. It is fabricated using SLE and metalized in the same way as for the trap with a similar trench design for electric isolation. The colored circuit paths in \cref{fig:assembly_multi_angle} show how each electrode is routed across the trap and circuit. At the bottom of the intermediate circuit, screws are inserted to clamp on a wire (\cref{fig:wire_circuit}) which is then connected to an external source through a feedthrough. On the trap, each electrode path terminates at a pad near their respective corner. These pads are aligned to their protruding counterpart on the intermediate circuit (\cref{fig:circuit_dimension}). The trap and the intermediate circuit are connected by physical contact at these pads. Poor contact between the pads may result in the desired signals not being delivered to the electrodes. Thus, the trap assembly is secured by a metal cap with four screws as shown in \cref{fig:trap_clamped}. High-voltage $1\,$nF shunt capacitors are also added to the dc electrode paths by conductive vacuum-compatible epoxy to suppress rf noise.

We stacked a pair of linear nanopositioners to move the cavity substrate in the $xy$ plane. A PEEK adapter is located below the nanopositioner stacks to achieve thermal isolation between the cavity and the rest of the system.

\begin{figure}[htbp]
    \centering
    \def\svgwidth{7cm}
\begingroup%
  \makeatletter%
  \providecommand\color[2][]{%
    \errmessage{(Inkscape) Color is used for the text in Inkscape, but the package 'color.sty' is not loaded}%
    \renewcommand\color[2][]{}%
  }%
  \providecommand\transparent[1]{%
    \errmessage{(Inkscape) Transparency is used (non-zero) for the text in Inkscape, but the package 'transparent.sty' is not loaded}%
    \renewcommand\transparent[1]{}%
  }%
  \providecommand\rotatebox[2]{#2}%
  \newcommand*\fsize{\dimexpr\f@size pt\relax}%
  \newcommand*\lineheight[1]{\fontsize{\fsize}{#1\fsize}\selectfont}%
  \ifx\svgwidth\undefined%
    \setlength{\unitlength}{699.32822881bp}%
    \ifx\svgscale\undefined%
      \relax%
    \else%
      \setlength{\unitlength}{\unitlength * \real{\svgscale}}%
    \fi%
  \else%
    \setlength{\unitlength}{\svgwidth}%
  \fi%
  \global\let\svgwidth\undefined%
  \global\let\svgscale\undefined%
  \makeatother%
  \begin{picture}(1,0.810712)%
    \lineheight{1}%
    \setlength\tabcolsep{0pt}%
    \put(0,0){\includegraphics[width=\unitlength,page=1]{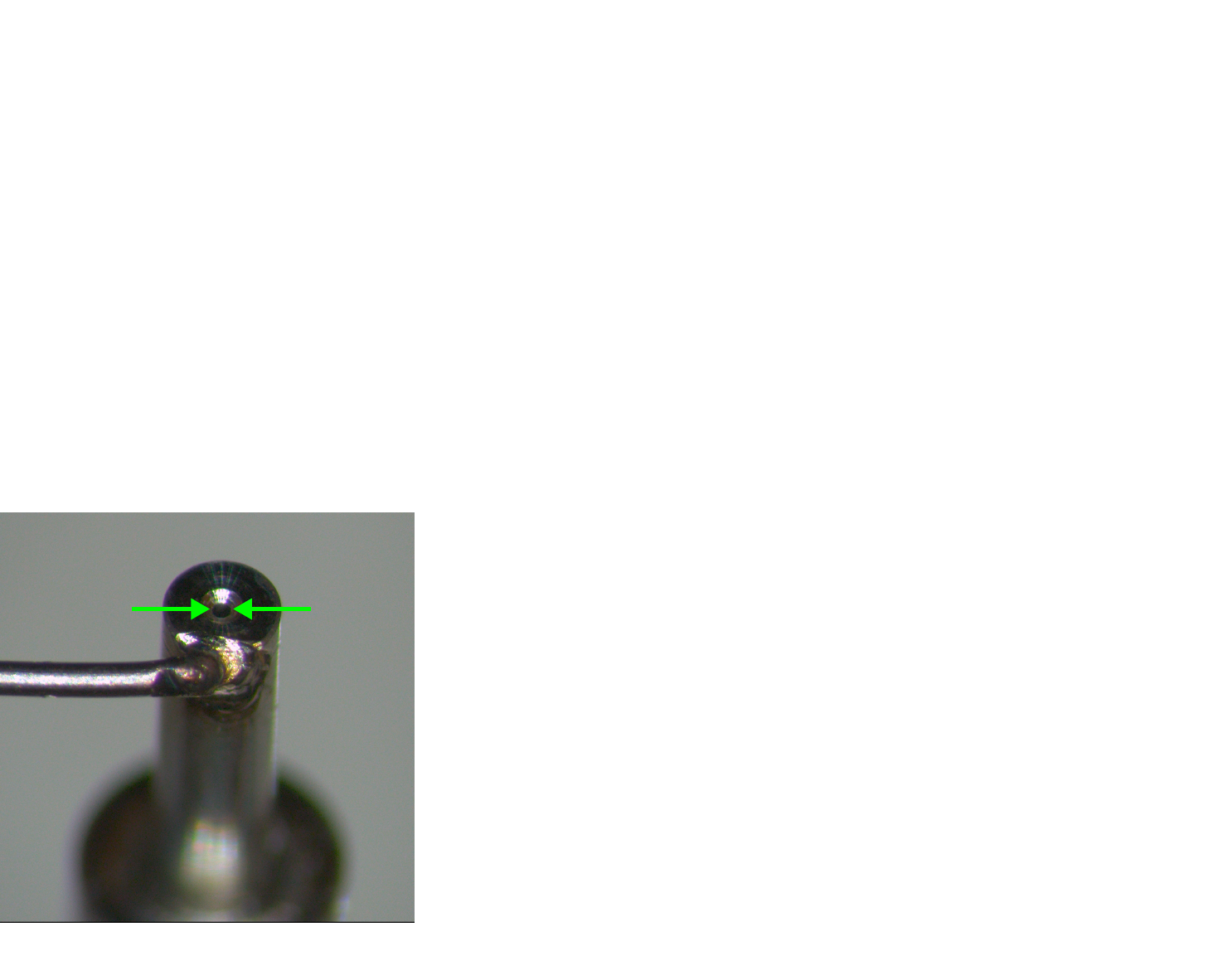}}%
    \put(0.11539123,0.35471583){\color[rgb]{0,1,0}\makebox(0,0)[lt]{\lineheight{1.25}\smash{\begin{tabular}[t]{l}0.2 mm\end{tabular}}}}%
    \put(0,0){\includegraphics[width=\unitlength,page=2]{oven.pdf}}%
    \put(0.43614939,0.13506877){\color[rgb]{0,0,0}\makebox(0,0)[lt]{\lineheight{1.25}\smash{\begin{tabular}[t]{l}Oven\end{tabular}}}}%
    \put(0.77979075,0.13469173){\color[rgb]{0,0,0}\makebox(0,0)[lt]{\lineheight{1.25}\smash{\begin{tabular}[t]{l}Collimator\end{tabular}}}}%
    \put(0.4307023,0.36059359){\color[rgb]{0,0,0}\makebox(0,0)[lt]{\lineheight{1.25}\smash{\begin{tabular}[t]{l}Shutter\end{tabular}}}}%
    \put(0,0){\includegraphics[width=\unitlength,page=3]{oven.pdf}}%
  \end{picture}%
\endgroup%

    \caption{Resistively heated oven source. Top: the oven schematics. The oven is machined with stainless steel with a small aperture of $0.2\,$mm to limit the divergence of the atomic flux. Bottom left: tantalum wire is spot welded to the oven to act as the heat source when passing the current. Bottom right: an additional collimator is added to collimate the atomic beam to the trap center.}
    \label{fig:oven}
\end{figure}

Two resistively heated ovens are included in our setup for generating atomic beams in the system. The ovens are machined from stainless steel into a hollow tube with a small opening and a threaded cap as shown in \cref{fig:oven}. Fresh calcium powder is loaded from the wider end before sealing with the screw cap. Spot welded tantalum wire provides necessary heating during the operation. A small opening combined with a collimator on the mount reduces the divergence and unnecessary deposition on the trap electrodes. Upon setting up the system, it is typical to ``crack'' the oven by by heating it to a high temperature to eject the impurities on the surface of the calcium pellets. This, however, would lead to unwanted deposition on the trap. So, we introduce a shutter that remained closed during the cracking process and can be opened using an actuator mounted on the nanopositioner (see \cref{fig:integration_overview}).

\section{Results and discussion}
\label{sec:res}
\begin{figure*}[htbp]
    \centering    

    \hfill
    \begin{subfigure}{0.45\linewidth}
    \includegraphics[width=0.9\textwidth]{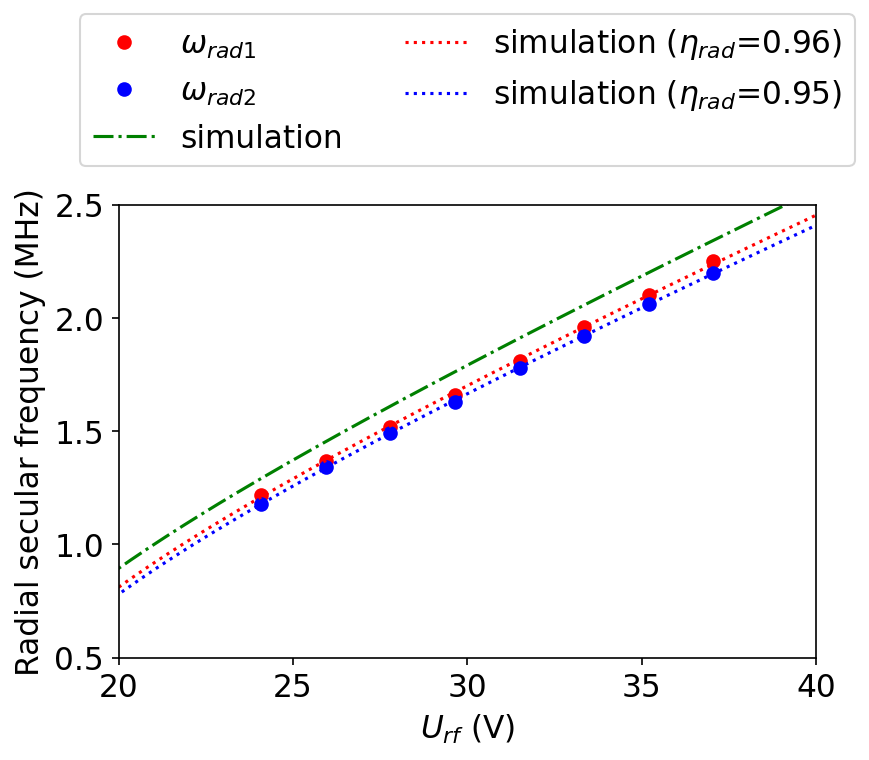}
    \caption{}
    \label{fig:radial_sec}    
    \end{subfigure}
    \hfill
    \begin{subfigure}{0.45\linewidth}
    \includegraphics[width=0.9\textwidth]{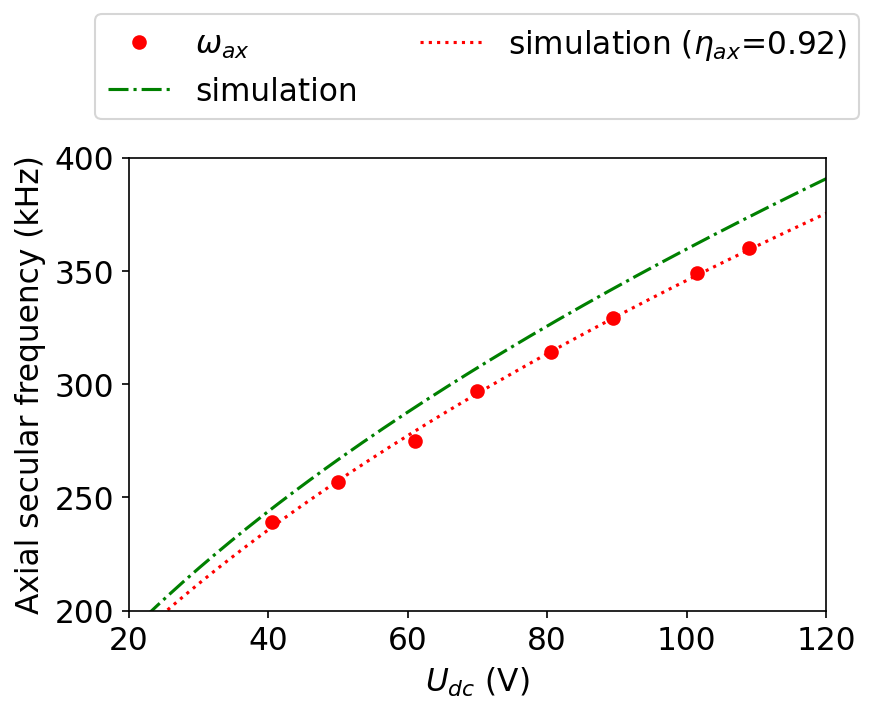}
    \caption{}
    \label{fig:axial_sec}    
    \end{subfigure}
    \hfill

    \caption{The \subref{fig:radial_sec} radial and \subref{fig:axial_sec} axial secular frequencies. The $\omega_{\text{rad1}}$ ($\omega_{\text{rad2}}$, $\omega_{\text{ax}}$) is in the direction spanned by the \texttt{rf gnd} (\texttt{rf}, \texttt{dc}) electrodes (see \cref{fig:simul_trap}). An offset is added to the simulation's secular frequency due to experiment condition to compare the simulation and the experimentally obtained data. A scaling factor $\eta_\text{rad}$ or $\eta_\text{ax}$ is applied to the simulation results to quantitatively determine the imperfection in the trap geometry. See main text for more information.}
    \label{fig:sec_freq}
\end{figure*}

We were able to trap single $\,^{40}\text{Ca}^+$ ions (\cref{fig:simul_trap}) doppler-cooled with $397\,$nm lasers and an $866\,$nm repumper. 
Magnetic field is applied along the $z$ axis to destabilize the dark states \cite{Berkeland2002DestabilizationSystems}.
Excess micromotion \cite{Berkeland1998MinimizationTrap} is compensated by applying a differential voltage along the \texttt{rf} electrodes or \texttt{rf gnd} electrodes (see \cref{fig:simul_trap}) to move the ion on the radial plane. We have confirmed that the ion's micromotion can be compensated along the trap axis over $100\,\mu$m from the trap center and observed no additional uncompensated excess micromotion. This implies that there is no residual rf field along the trap axis originating from the misalignment or imperfections of the trap fabrication. 

To characterize the trapping potential of our trap, we measure the secular frequencies ($\omega_{\text{rad1}}$, $\omega_{\text{rad2}}$, $\omega_{\text{ax}}$) along the \texttt{rf}, \texttt{rf gnd} and \texttt{dc} electrodes respectively with the results presented in \cref{fig:sec_freq}. The ions' secular frequency is obtained by applying a small tickling rf signal on the electrode which will preferentially excite the ion motion along the direction defined by the electrode. The frequency of the tickling rf is scanned at a sufficiently small step size. When the tickling rf frequency is resonant with the secular frequency, the ion's motion is excited and can be confirmed by the ion's motion on the camera or a drop in a Photomultiplier tube (PMT) photon count \cite{Carette1997ProbingTrap}. 

We use the simulated trap's potential to determine the fabrication quality with the simulation details provided in appendix \ref{app:sim}. However, the overall trap potential in the real system might deviate from the ideal simulation due to fabrication imperfection or unwanted coupling and leakage. We consider a model for quantify the fabrication imperfection in our trap. The following equations are used to fit the simulated radial and axial secular frequency to measurements:
\begin{align}
\omega_{\text{fit, rad}}^2 &= \frac{\Omega_\text{rf}^2}{4} \left[ \frac{(\eta_\text{rad}q)^2}{2}+a_\text{rad}+b_\text{rad} \right],
\label{eq:rad_fit_eq} \\
\omega_{\text{fit, ax}}^2 &= \frac{\Omega_\text{rf}^2}{4}(\eta_\text{ax}a_\text{ax}+b_\text{ax}),
\label{eq:ax_fit_eq}
\end{align}
where $\Omega_\text{rf}$ is the driving frequency. $\eta_\text{rad}$ and $\eta_\text{ax}$ are imperfection coefficients in the real system that changes the dependency to $U_\text{rf}$ and dc voltages $U_\text{dc}$ respectively. A deviation from unity for $\eta_\text{rad}$ ($\eta_\text{ax}$) implies the presence of imperfection for the rf (dc) electrodes. On the other hand, additional confinements due to other sources are captured in the constants $b_\text{rad}$ and $b_\text{ax}$, which are determined experimentally. The $q$ and $a$ parameters are obtained through the simulation.

The efficiency of the voltage delivered to the trap electrodes can affect the coefficients $\eta_\text{rad}$ and $\eta_\text{ax}$, so we take the following measures to decouple the influence. For $\eta_\text{ax}$, we calibrate the dc voltage such that the final measured voltage before the vacuum chamber is the same as the set voltage. For the rf voltage delivery, we used a helical resonator \cite{Siverns2012OnResonators} which filters and amplifies the input voltage. Direct measurement of such a system in isolation is inappropriate without consideration for the impedance matching condition at the trap. Methods for precisely determining the actual rf power at the trap require prior inclusion of a matching circuit \cite{Gandolfi2012CompactTraps, Johnson2016ActivePotentials, Park2021AResonator}. In our setup, we instead consider a simple estimation of the voltage at the trap by estimating the amplification factor $A$ as
\begin{equation}
    A = \frac{U_\text{rf}}{U_\text{in}} \approx \frac{\kappa \sqrt{2PQ}}{U_\text{in}},
    \label{eq:amp}
\end{equation}
where
\begin{equation}
    \kappa=\sqrt{QR}.
\end{equation}
The helical resonator amplifies the input voltage $U_\text{in}$ to the voltage at the trap $U_\text{rf}$. The amplification is related to the input power $P$, the resistance $R$ of the load and connections, and the quality factor $Q$ of the system. The $Q$ of the entire system is obtained from a network analyzer, while R is measured at 5 MHz using a LCR meter. The measured values are $Q=82.91\pm0.83$ and $R=5\pm2$, with the calculated amplification as $A=26.20\pm5.25$.

By including the necessary corrections presented in \cref{eq:rad_fit_eq} and \cref{eq:ax_fit_eq}, we compare the simulation and experimental data in \cref{fig:sec_freq}. Constant offsets $b_\text{rad}$ and $b_\text{ax}$ were added to the simulation results presented. We experimentally verify that the source of the $b_\text{rad}$ and $b_\text{ax}$ is independent of our applied voltage $U_\text{rf}$ and $U_\text{dc}$ and remained constant throughout the period of data collection. The difference in the values of $b_\text{rad}$ between $\omega_{\text{rad1}}$ and $\omega_{\text{rad2}}$ is negligeble at only $0.58\%$. Thus, the $b_\text{rad}$ used is taken to be the average. Deviation between the simulation and experiment results without the scaling factors $\eta_\text{rad}$ and $\eta_\text{ax}$ was observed and indicates imperfection in our trap. From the fitting, the degrees of imperfection for ($\omega_{\text{rad1}}$, $\omega_{\text{rad2}}$, $\omega_{\text{ax}}$) are $(0.96, 0.95, 0.92)$. Due to our simple estimation of the amplification factor in \cref{eq:amp}, the uncertainty in the $\eta_\text{rad}$ obtained for $\omega_{\text{rad1}}$ and $\omega_{\text{rad2}}$ is $20\%$. In all cases, the $\eta_k<1$ indicates that the measured confinement is weaker than the simulation, which can be attributed to either receded electrodes due to the SLE fabrication (see appendix \ref{app:caveats}) or unknown voltage source leaking into the system. Nevertheless, the values are still close to unity, which implies a good fabrication quality. The imperfection coefficients can be compared directly to the values obtained for a monolithic trap fabricated with laser cutting \cite{Wang2020CoherentlyTrap}. The values obtained with our trap are closer to unity, suggesting an improved machining precision with SLE processing.

\section{Conclusion}
In this work, we presented a novel monolithic linear trap for integrating a microcavity that will allow the realization of an efficient ion-cavity coupling in a linear ion trap. Our trap features high optical access, while miniaturization provides suitable shielding for integrating a microcavity. Furthermore, we have demonstrated the trapping of ions using our novel monolithic linear trap. From the measurement of secular frequencies, we show that the fabrication technique yields a trap with low imperfection. In future works, the trap can be further miniaturized and packaged with a printed circuit board (PCB) for compactness. Furthermore, additional dc electrodes can be added to the rf electrodes to facilitate the shuttling of the ions in and out of the cavity interaction zone and to conduct localized excess micromotion compensation. In addition, we plan to integrate the cavity substrate with the machined fibers in our trap to demonstrate efficient ion-cavity coupling. With tuned parameters, developing an ion trap with the SLE technique is highly repeatable with good tolerance. Mass fabrication of state-of-the-art ion traps compatible with microcavities will enable interesting real-world applications in many fields, such as quantum information processing and quantum metrology.

\section{Acknowledgements}
We thank Joel Morley, Chitose Maruko, Alexander Henry Hodges, and Artem Podlesnyy for their help and support. This work was supported by the JST Moonshot R\textnormal{\&}D (Grant Number JPMJMS2063) and MEXT Quantum Leap Flagship Program (Grant Number JPMXS0118067477). The authors also acknowledge financial support from the OIST Graduate University, in particular the OIST Proof of Concept Program - Seed Phase Project (R$11\_59$).

\appendix

\section{Simulation}
\label{app:sim}

\begin{figure*}[htbp]
    \centering
    \begin{subfigure}{0.3\linewidth}
    \includegraphics[width=0.9\textwidth]{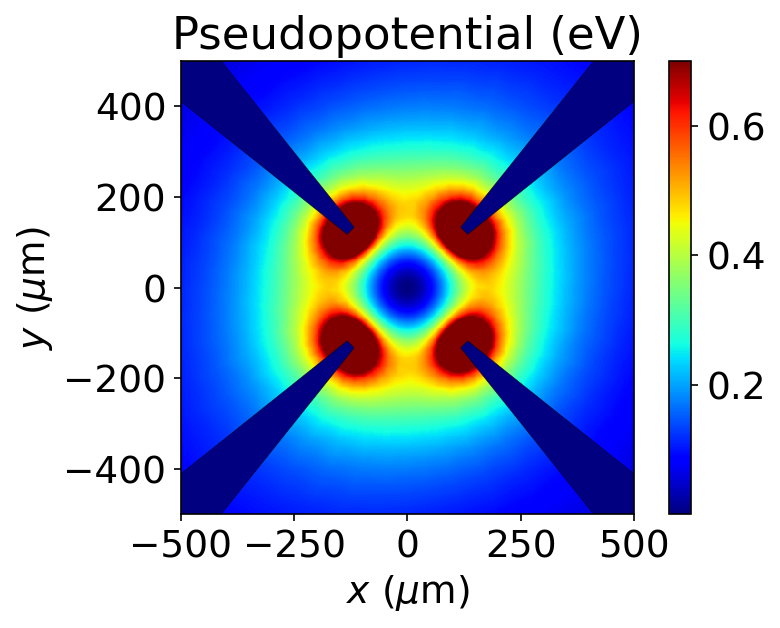}
    \caption{}
    \label{fig:potential}    
    \end{subfigure}
    \begin{subfigure}{0.3\linewidth}
    \includegraphics[width=0.9\textwidth]{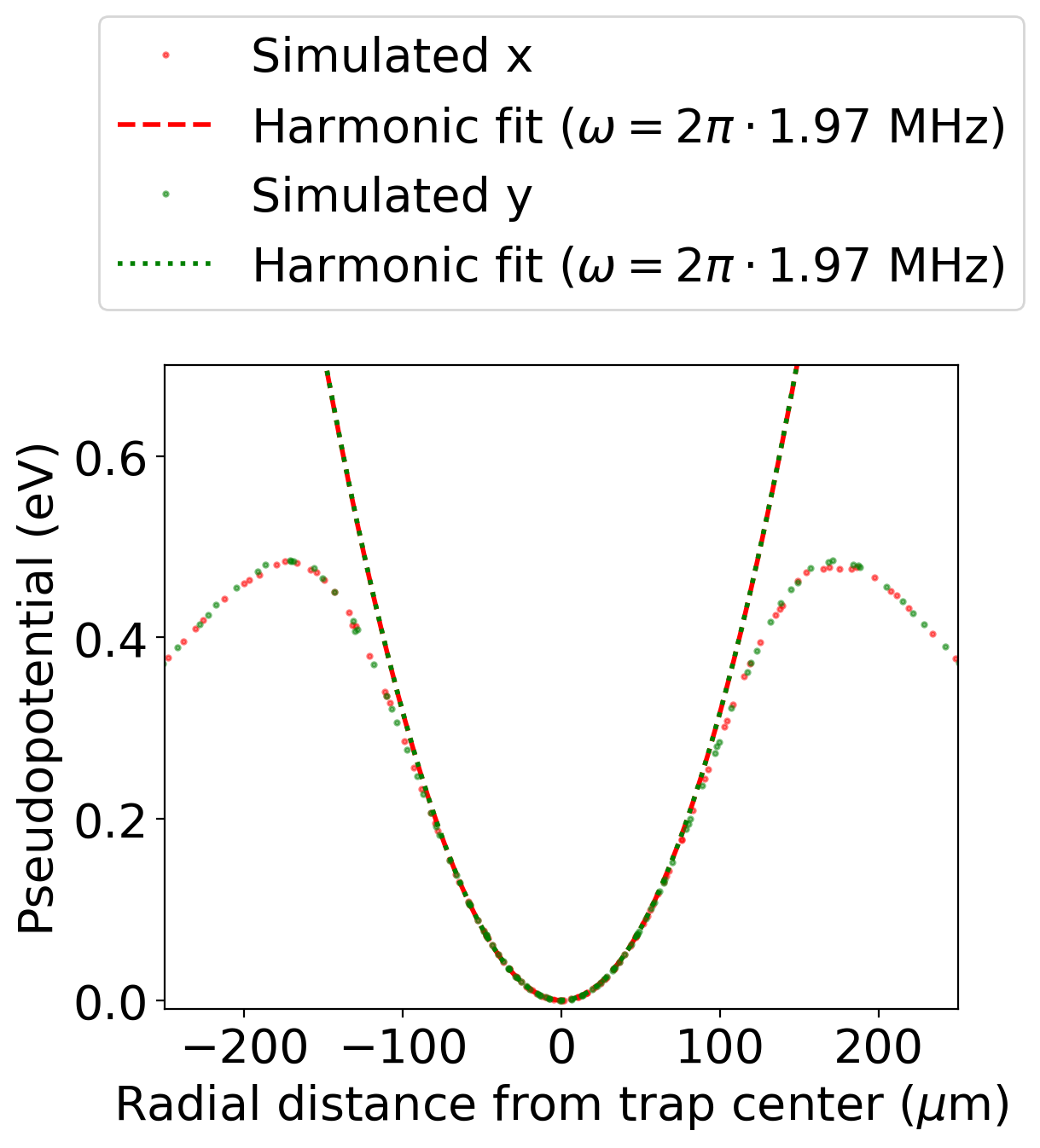}
    \caption{}
    \label{fig:radial_pot_no_pyr}    
    \end{subfigure}
    \begin{subfigure}{0.3\linewidth}
    \includegraphics[width=0.9\textwidth]{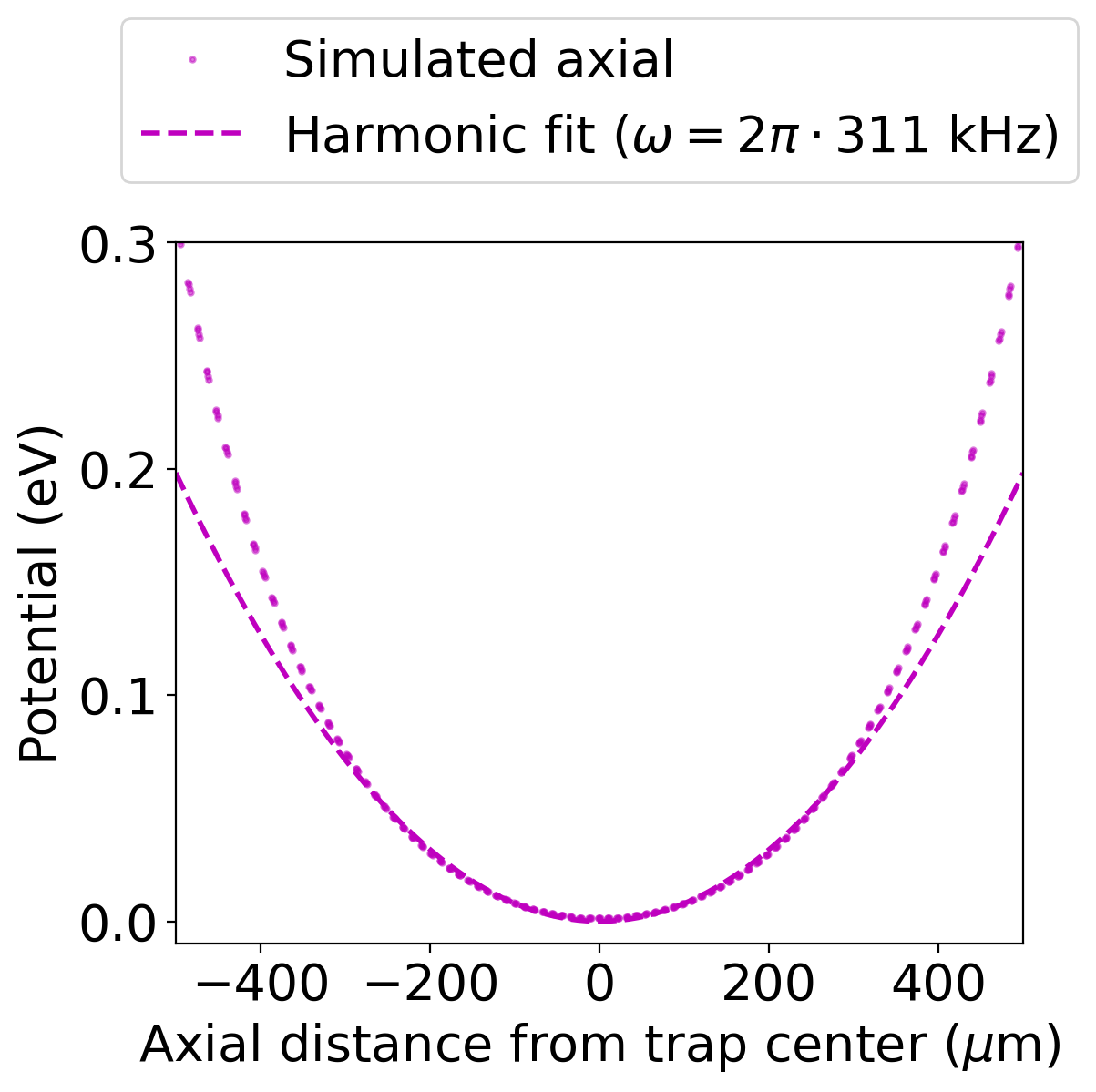}
    \caption{}
    \label{fig:axial_pot}    
    \end{subfigure}
    
    \caption{The trap potential. \subref{fig:potential} The 2D cross-section of the trap's pseudopotential in the radial plane. The distance between the trap center and electrode is $177\,\mu$m. The pseudopotential has translational symmetry along the trap ($z$) axis. \subref{fig:radial_pot_no_pyr} The radial pseudopotential of the trap is fitted with a quadratic function. The potential forms a cylindrically symmetrical harmonic potential about the rf-null close to the center with the secular frequencies $(\omega_\text{x}, \omega_\text{y})=2\pi\times(1.97, 1.97)\,$MHz. The applied rf voltage has a peak voltage of $30\,$V and frequency $2\pi\times15.82$MHz. \subref{fig:axial_pot} The axial secular frequencies is $2\pi\times 311\,$kHz when $60\,$V is applied on the endcap electrodes.}
    \label{fig:pseudopotential}
\end{figure*}

We simulate the trapping capability with the trap geometry presented in \cref{fig:simul_trap} via finite element analysis with COMSOL Multiphysics. The simulation here takes into account the shorted endcap mentioned in the main text. The radial trapping potential is shown in \cref{fig:potential} with the simulation condition given in \cref{tab:trap_parameter}. The simulated potentials are well approximated by harmonic potentials near the trap center. The radial secular frequencies of our trap are $(\omega_\text{x},\ \omega_\text{y})=2\pi\times(1.97,\ 1.97)\,$MHz with trap depth of $0.48\,$eV. Due to the symmetry of our trap, the radial secular frequencies have rotational symmetry near the trap center. On the other hand, the axial secular frequency is $2\pi\times 311\,$kHz. To consider the effect of charging on the cavity substrate, we assumed the values of surface charge densities $\sigma_\text{s}=+50\,$e$\,\mu\text{m}^{-2}$ as obtained in an experiment involving optical fiber-based cavities \cite{KumarVerma2020ProbingIon}. By superposing compensation dc on the electrodes, we require around $1\,$V to overcome the shift in the potential. Thus, our simulation results suggest that our trap design is capable of trapping ions and correcting for stray charges. A more comprehensive numerical study on cavity integration is presented in \cite{EzraSimulationPaper}.

\begin{table}
    \centering
    \begin{tabular}{cc}
        Parameter & Value \\ [0.5ex] 
        \hline\hline
        Ion species & $\,^{40}\text{Ca}^+$ \\
        rf freq. $\Omega_\text{RF}$ & $2\pi\times15.82\,$MHz \\
        rf peak voltage $V_\text{RF}$ & $30\,$V\\
        Endcap voltage $V_\text{DC}$ & $60\,$V \\
        Trap center to electrode distance & $177\,\mu$m \\
        Trap material & Gold
    \end{tabular}
    \caption{Trap driving parameters used for the finite-element simulations.}
    \label{tab:trap_parameter}
\end{table}

\section{Fabrication specifications}
\label{app:fab_specs}
In our fabrication setup, we use UV fused silica (Optostar Ltd, Japan) as the substrate and etched in potassium hydroxide (KOH) after laser writing. Our in-house production utilizes LightFab 3D Printer (LightFab GmbH, Germany) for laser writing. The laser parameters and etching procedure were configured to produce an optimized selectivity of $>1000$ for fused silica \cite{Gottmann2017SelectiveSpeed}. The Satsuma (Amplitude Systemes, France) laser source produces laser pulse at $1030\,$nm wavelength, with $750\,$kHz repetition rate, and $200\,$mm/s writing velocity. The pulse duration is set to $1000\,$fs, and a pulse energy of about $500\,$nJ, where the selectivity remained robust to parameter fluctuation \cite{Gottmann2017SelectiveSpeed}. We clean the fused silica with acetone and isopropyl alcohol before and after the laser writing. Then, etching is done by submerging the substrate in KOH over the course of 48 hours. The KOH solution is kept at $85\,^\circ$C and agitated in an ultrasonic bath.
\begin{figure}[htbp]
    \centering
    
    \begin{subfigure}{0.9\linewidth}
    \centering
    \def\svgwidth{8cm}
    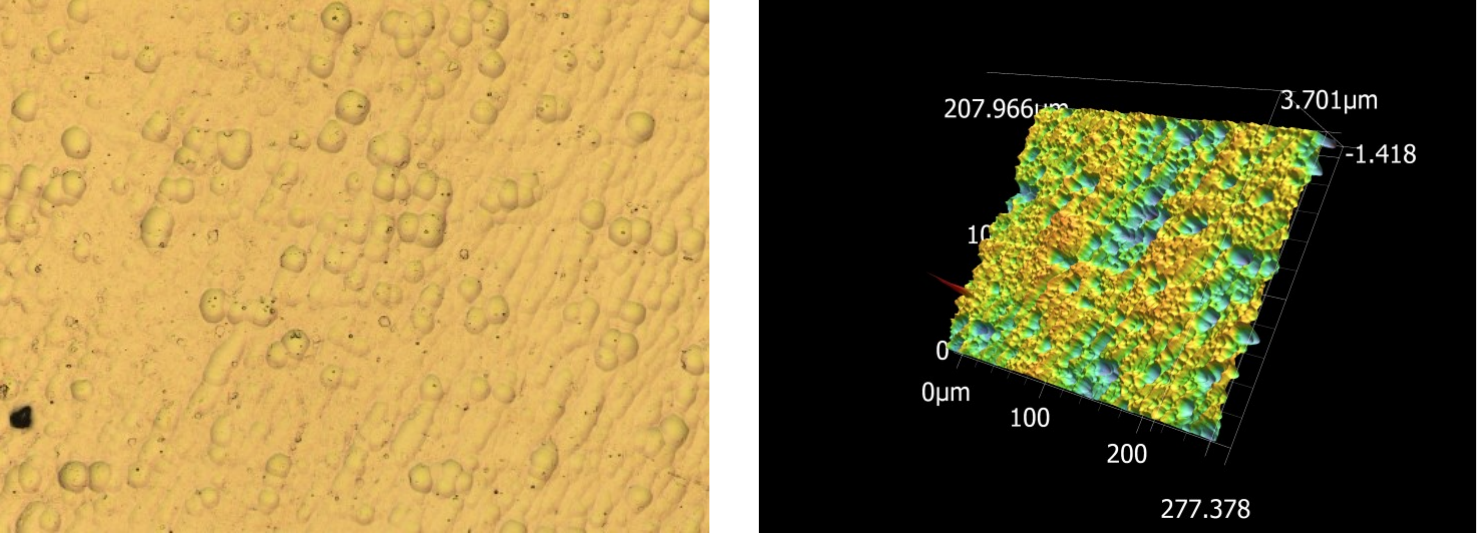
    \caption{}
    \label{fig:surface_roughness_femtoprint}
    \end{subfigure}

    \begin{subfigure}{0.9\linewidth}
    \centering
    \vspace{2mm}
    \def\svgwidth{8cm}
    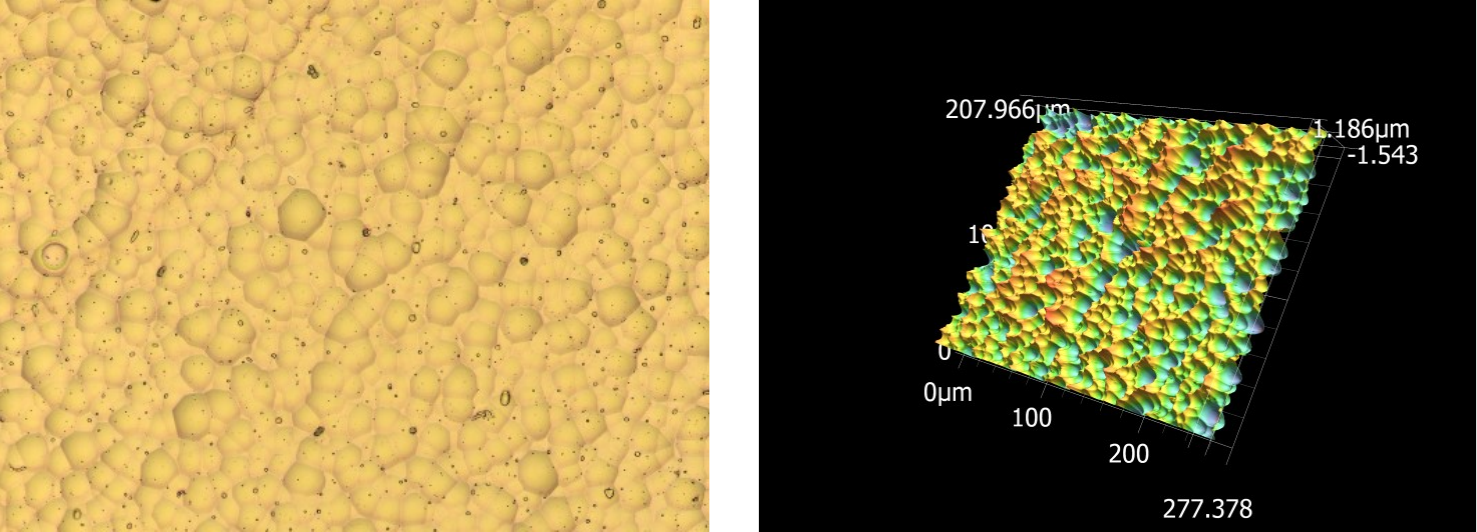
    \caption{}
    \label{fig:surface_roughness_inhouse}
    \end{subfigure}

    \caption{The surface roughness on the electrodes produced with SLE by \subref{fig:surface_roughness_femtoprint} FEMTOprint and \subref{fig:surface_roughness_inhouse} in-house. On the left is image taken by microscope at 50x magnification, and the roughness is measured by a laser microscope on the right. The measured roughness $Ra$ is $216\,$nm and $281\,$nm for FEMTOprint and in-house trap respectively for an area of $200\times200\,\mu$m.}
    \label{fig:surface_roughness}
\end{figure}

In the ion trapping experiment, the trap used was manufactured by FEMTOprint SA. Despite our ability to fabricate the trap in-house, the decision to use the trap provided by FEMTOprint is due to the better surface quality as shown in \cref{fig:surface_roughness}. The surface roughness measured over an area of $200\times200\,\mu$m for FEMTOprint is $R_a=216\pm1\,$nm, marginally better than our in-house capability at $R_a=281\pm2\,$nm. Other prints, such as oven mount, shutter, and intermediate circuit, were fabricated in-house. After the assembly, we noticed that one of the endcap electrodes was shorted to the ground due to a worn-out Kapton wire. We ran a simulation with a grounded dc endcap electrode and verified that the deviation only resulted in a small shift of the rf-null line from the ideal case.

\section{Fiber insertion into the cavity substrate}
\label{app:cavity_substrate}
\begin{figure}[!htbp]
    \def\svgwidth{7cm}
\begingroup%
  \makeatletter%
  \providecommand\color[2][]{%
    \errmessage{(Inkscape) Color is used for the text in Inkscape, but the package 'color.sty' is not loaded}%
    \renewcommand\color[2][]{}%
  }%
  \providecommand\transparent[1]{%
    \errmessage{(Inkscape) Transparency is used (non-zero) for the text in Inkscape, but the package 'transparent.sty' is not loaded}%
    \renewcommand\transparent[1]{}%
  }%
  \providecommand\rotatebox[2]{#2}%
  \newcommand*\fsize{\dimexpr\f@size pt\relax}%
  \newcommand*\lineheight[1]{\fontsize{\fsize}{#1\fsize}\selectfont}%
  \ifx\svgwidth\undefined%
    \setlength{\unitlength}{186.13471805bp}%
    \ifx\svgscale\undefined%
      \relax%
    \else%
      \setlength{\unitlength}{\unitlength * \real{\svgscale}}%
    \fi%
  \else%
    \setlength{\unitlength}{\svgwidth}%
  \fi%
  \global\let\svgwidth\undefined%
  \global\let\svgscale\undefined%
  \makeatother%
  \begin{picture}(1,0.74768521)%
    \lineheight{1}%
    \setlength\tabcolsep{0pt}%
    \put(0,0){\includegraphics[width=\unitlength,page=1]{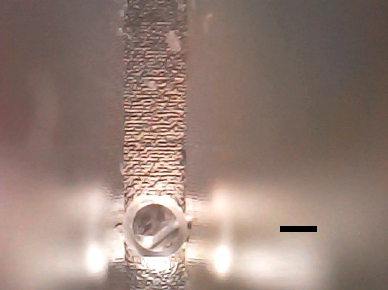}}%
    \put(0.69135376,0.07347312){\color[rgb]{0,0,0}\makebox(0,0)[lt]{\lineheight{1.25}\smash{\begin{tabular}[t]{l}0.1 mm\end{tabular}}}}%
    \put(0,0){\includegraphics[width=\unitlength,page=2]{cavity_assembly.pdf}}%
    \put(0.64051568,0.59032567){\color[rgb]{0,1,0}\makebox(0,0)[lt]{\lineheight{1.25}\smash{\begin{tabular}[t]{l}substrate \\tip\end{tabular}}}}%
    \put(0.7129178,0.36416447){\color[rgb]{0,1,0}\makebox(0,0)[lt]{\lineheight{1.25}\smash{\begin{tabular}[t]{l}fiber\\facet\end{tabular}}}}%
  \end{picture}%
\endgroup%

    \vspace{1mm}
    
    \caption{A stripped coreless fiber is inserted a cavity substrate. The fiber diameter is 125~$\mu$m.}
    \label{fig:cavity_assembly}
\end{figure}

\begin{figure*}[!ht]
    \centering

    \begin{subfigure}{0.9\textwidth}
    \centering
    \hfill
    \def\svgwidth{4cm}
\begingroup%
  \makeatletter%
  \providecommand\color[2][]{%
    \errmessage{(Inkscape) Color is used for the text in Inkscape, but the package 'color.sty' is not loaded}%
    \renewcommand\color[2][]{}%
  }%
  \providecommand\transparent[1]{%
    \errmessage{(Inkscape) Transparency is used (non-zero) for the text in Inkscape, but the package 'transparent.sty' is not loaded}%
    \renewcommand\transparent[1]{}%
  }%
  \providecommand\rotatebox[2]{#2}%
  \newcommand*\fsize{\dimexpr\f@size pt\relax}%
  \newcommand*\lineheight[1]{\fontsize{\fsize}{#1\fsize}\selectfont}%
  \ifx\svgwidth\undefined%
    \setlength{\unitlength}{139.91331085bp}%
    \ifx\svgscale\undefined%
      \relax%
    \else%
      \setlength{\unitlength}{\unitlength * \real{\svgscale}}%
    \fi%
  \else%
    \setlength{\unitlength}{\svgwidth}%
  \fi%
  \global\let\svgwidth\undefined%
  \global\let\svgscale\undefined%
  \makeatother%
  \begin{picture}(1,1.11940567)%
    \lineheight{1}%
    \setlength\tabcolsep{0pt}%
    \put(0,0){\includegraphics[width=\unitlength,page=1]{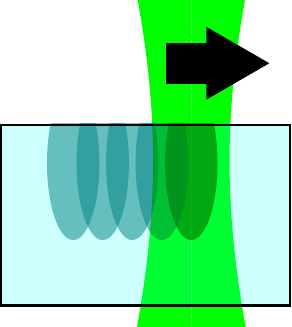}}%
    \put(0.16474967,0.92488753){\color[rgb]{0,0,0}\makebox(0,0)[lt]{\lineheight{1.25}\smash{\begin{tabular}[t]{l}Scanning \\direction\end{tabular}}}}%
    \put(0,0){\includegraphics[width=\unitlength,page=2]{sle_caveat_beam_shape.pdf}}%
    \put(0.22015541,0.11744515){\color[rgb]{0,0,0}\makebox(0,0)[lt]{\lineheight{1.25}\smash{\begin{tabular}[t]{l}x\end{tabular}}}}%
    \put(0.03221113,0.32380973){\color[rgb]{0,0,0}\makebox(0,0)[lt]{\lineheight{1.25}\smash{\begin{tabular}[t]{l}z\end{tabular}}}}%
    \put(0.4953934,0.12171798){\color[rgb]{0,0,0}\makebox(0,0)[lt]{\lineheight{1.25}\smash{\begin{tabular}[t]{l}scallop\end{tabular}}}}%
    \put(0,0){\includegraphics[width=\unitlength,page=3]{sle_caveat_beam_shape.pdf}}%
  \end{picture}%
\endgroup%

    \hfill
    \def\svgwidth{8cm}
    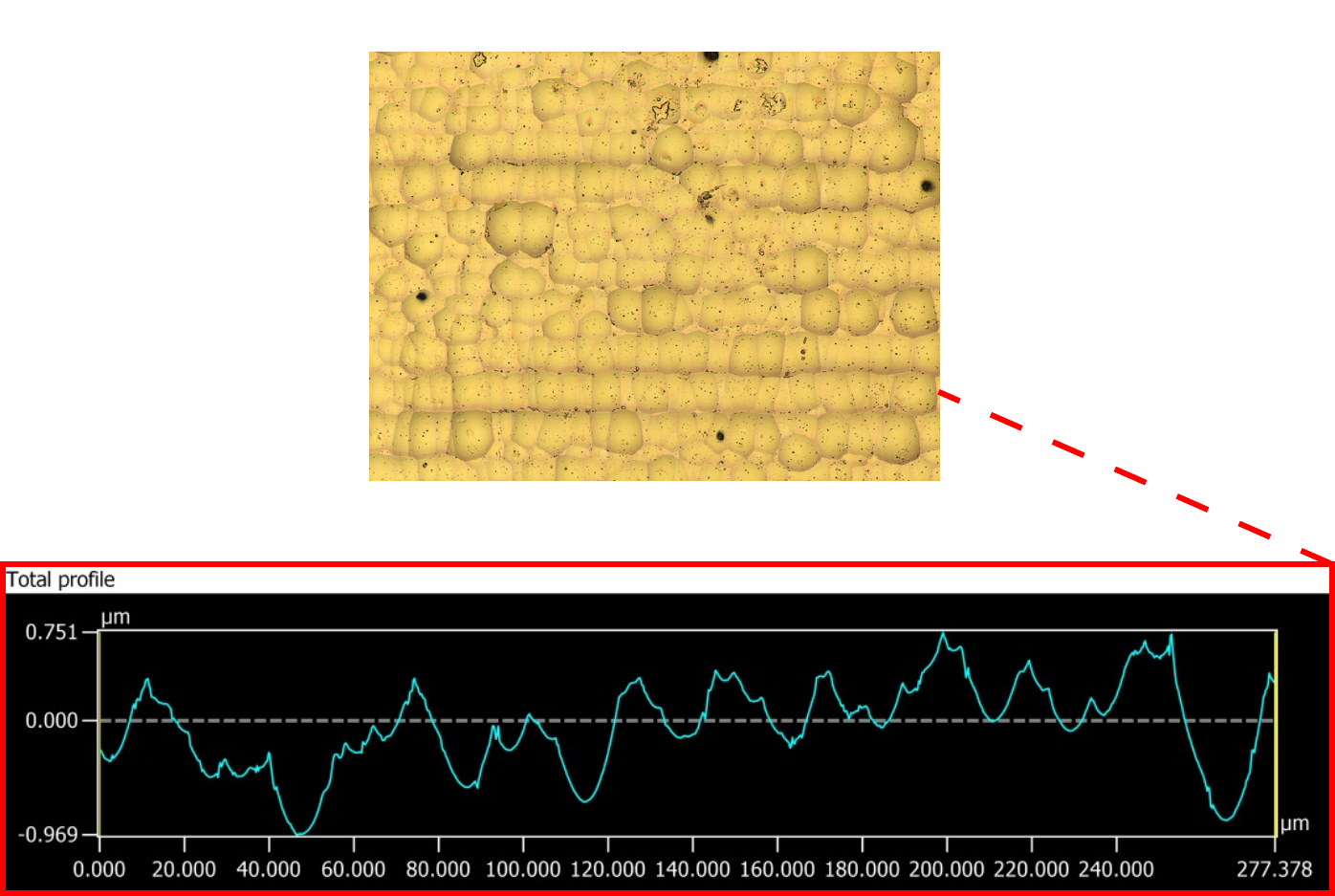
    \hfill
    \vspace*{0.1cm}
    
    \caption{}
    \label{fig:sle_caveat_beam_shape}
    \vspace*{0.5cm}
    \end{subfigure}

    \hfill
    \begin{subfigure}{0.3\linewidth}
    \centering
    \def\svgwidth{4.5cm}
\begingroup%
  \makeatletter%
  \providecommand\color[2][]{%
    \errmessage{(Inkscape) Color is used for the text in Inkscape, but the package 'color.sty' is not loaded}%
    \renewcommand\color[2][]{}%
  }%
  \providecommand\transparent[1]{%
    \errmessage{(Inkscape) Transparency is used (non-zero) for the text in Inkscape, but the package 'transparent.sty' is not loaded}%
    \renewcommand\transparent[1]{}%
  }%
  \providecommand\rotatebox[2]{#2}%
  \newcommand*\fsize{\dimexpr\f@size pt\relax}%
  \newcommand*\lineheight[1]{\fontsize{\fsize}{#1\fsize}\selectfont}%
  \ifx\svgwidth\undefined%
    \setlength{\unitlength}{423.29474502bp}%
    \ifx\svgscale\undefined%
      \relax%
    \else%
      \setlength{\unitlength}{\unitlength * \real{\svgscale}}%
    \fi%
  \else%
    \setlength{\unitlength}{\svgwidth}%
  \fi%
  \global\let\svgwidth\undefined%
  \global\let\svgscale\undefined%
  \makeatother%
  \begin{picture}(1,0.84998152)%
    \lineheight{1}%
    \setlength\tabcolsep{0pt}%
    \put(0,0){\includegraphics[width=\unitlength,page=1]{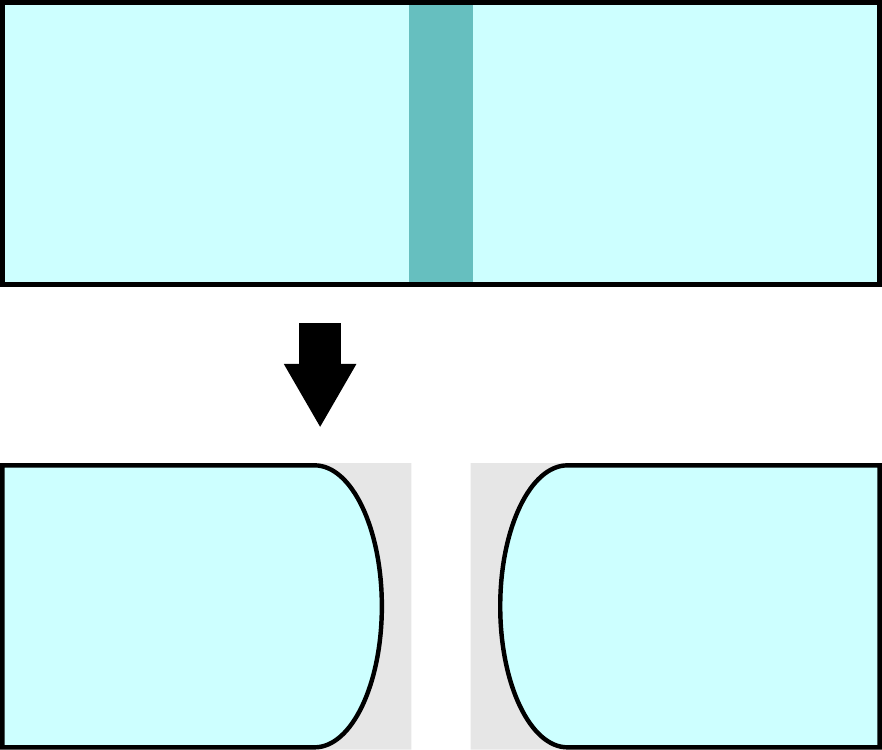}}%
    \put(0.42163792,0.40513734){\color[rgb]{0,0,0}\makebox(0,0)[lt]{\lineheight{1.25}\smash{\begin{tabular}[t]{l}Etching\end{tabular}}}}%
    \put(0,0){\includegraphics[width=\unitlength,page=2]{sle_caveat_overetching.pdf}}%
    \put(0.57610564,0.71964052){\color[rgb]{0,0,0}\makebox(0,0)[lt]{\lineheight{1.25}\smash{\begin{tabular}[t]{l}modified\\material\end{tabular}}}}%
    \put(0,0){\includegraphics[width=\unitlength,page=3]{sle_caveat_overetching.pdf}}%
    \put(0.6254025,0.10347281){\color[rgb]{0,0,0}\makebox(0,0)[lt]{\lineheight{1.25}\smash{\begin{tabular}[t]{l}over-etching\end{tabular}}}}%
  \end{picture}%
\endgroup%

    \caption{}
    \label{fig:sle_caveat_overetching}
    \end{subfigure}
    \hfill
    \begin{subfigure}{0.4\linewidth}
    \centering
    \def\svgwidth{7cm}
\begingroup%
  \makeatletter%
  \providecommand\color[2][]{%
    \errmessage{(Inkscape) Color is used for the text in Inkscape, but the package 'color.sty' is not loaded}%
    \renewcommand\color[2][]{}%
  }%
  \providecommand\transparent[1]{%
    \errmessage{(Inkscape) Transparency is used (non-zero) for the text in Inkscape, but the package 'transparent.sty' is not loaded}%
    \renewcommand\transparent[1]{}%
  }%
  \providecommand\rotatebox[2]{#2}%
  \newcommand*\fsize{\dimexpr\f@size pt\relax}%
  \newcommand*\lineheight[1]{\fontsize{\fsize}{#1\fsize}\selectfont}%
  \ifx\svgwidth\undefined%
    \setlength{\unitlength}{381.38060562bp}%
    \ifx\svgscale\undefined%
      \relax%
    \else%
      \setlength{\unitlength}{\unitlength * \real{\svgscale}}%
    \fi%
  \else%
    \setlength{\unitlength}{\svgwidth}%
  \fi%
  \global\let\svgwidth\undefined%
  \global\let\svgscale\undefined%
  \makeatother%
  \begin{picture}(1,0.69123087)%
    \lineheight{1}%
    \setlength\tabcolsep{0pt}%
    \put(0,0){\includegraphics[width=\unitlength,page=1]{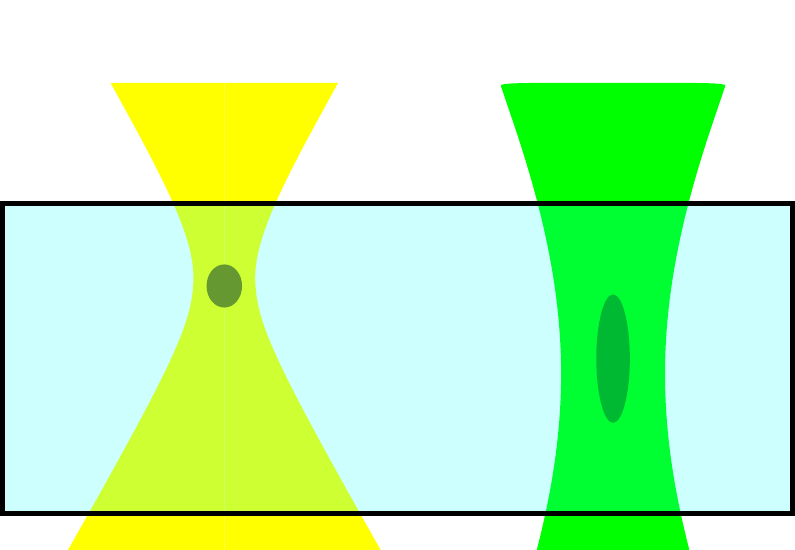}}%
    \put(0.62016949,0.63404008){\color[rgb]{0,0,0}\makebox(0,0)[lt]{\lineheight{1.25}\smash{\begin{tabular}[t]{l}Low NA\end{tabular}}}}%
    \put(0.11912139,0.6314667){\color[rgb]{0,0,0}\makebox(0,0)[lt]{\lineheight{1.25}\smash{\begin{tabular}[t]{l}High NA\end{tabular}}}}%
  \end{picture}%
\endgroup%

    \caption{}
    \label{fig:sle_caveat_depth_resolution}
    \end{subfigure}
    \hfill

    \caption{Limitations of using SLE for microfabrication. \subref{fig:sle_caveat_beam_shape} Formation of scallops due to the discrete shots of laser pulses. Left: schematics of the formation due to the elliptical LAZs. Right: a sample after SLE shows scallops of $\sim 1\,\mu$m forming along the scanning path in the lateral plane. The surface roughness $Ra$ is measured to be 546 nm. Note that the surface roughness shown in \cref{fig:surface_roughness} is measured on the plane along the laser beam. \subref{fig:sle_caveat_overetching} Degree of over-etching depends on the distance from the substrate surface. Scales are exaggerated. \subref{fig:sle_caveat_depth_resolution} Aspect ratio and maximum depth of the LAZ at different NA.}
    \label{fig:sle_limitation}
\end{figure*}
We fabricated the cavity substrate presented in \cref{fig:cavity_substrate} with SLE and constructed an assembly setup for inserting a fiber to a hollow channel of the substrate. Since the fiber will be machined into an FFPC with sensitive high-reflectivity coating, the assembly features cameras pointing from multiple angles to ensure smooth insertion of the fiber without collision with the channel's inner wall. A successfully inserted fiber inside the cavity substrate is shown in \cref{fig:cavity_assembly}. The fiber used for this demonstration is yet to be machined with $\text{CO}_2$ laser ablation.

\section{Caveats}
\label{app:caveats}
Despite the promise of higher-quality microfabrication with the SLE technique. We noted several important limitations during our fabrication. This section discusses the caveats of working with SLE.

\subsection{Surface roughness}
Due to the focusing of the laser pulse, the cross-section of the LAZ has an elliptical shape elongated in the beam direction and may not be desirable. \Cref{fig:sle_caveat_beam_shape} shows the effect of discrete steps of the elliptical pulses during the laser scanning that resulted in the formation of scallops \cite{Sugioka2003Three-dimensionalLaser, Ju2011FabricationMicromachining}. Beam shaping techniques such as slit \cite{Withford2005SlitGlasses, Diez-Blanco2007DeepWriting}, astigmatic \cite{Polli2003FemtosecondBeams}, and spatiotemporal beam shaping \cite{Xu2010FabricationPulses, Li2018Aberration-insensitivePulses} has demonstrated ability to produce LAZ of circular cross-section. Reduction of the step size between the LAZ can also reduce the size of the scallop at the cost of longer writing time.

We consider the effect of arithmetic surface roughness $Ra$ of the electrode in the limits of our observed values (\cref{fig:sle_caveat_beam_shape}): $Ra<1\,\mu$m and ion-electrode distance on the orders of $100\,\mu$m. The contribution to the asymmetry in the electric field in the trap center is minimal due to the fluctuation being averaged out at a large ratio of ion-electrode distance to $Ra$. Beyond the static behavior, the roughness of the surface may affect the motional heating of ions. Due to exposure to the atmosphere, the gold electrode surface can adsorb atomic impurities, which leads to the randomly distributed fluctuating dipoles on the surface \cite{Safavi-Naini2011MicroscopicTraps}. This adsorbate dipole fluctuations model was employed to evaluate the effect of surface roughness on the motional heating of ions \cite{Lin2016EffectsIons}. The anomalous heating is suppressed when the adsorbates are concentrated in the ``valley'', or a positive curvature region of the surface. Therefore the effect of the anomalous heating is reduced with increasing roughness in this model. In the case of the SLE process, since the surface features are composed of mostly positive curvatures, the heating rate may be reduced.

\subsection{Over-etching}
Selectivity is a metric for quantifying the etching quality,  defined as the ratio between the etching rate of the laser-modified and unmodified material. Ideally, the selectivity should be as high as possible, and our choice of KOH is motivated by the lack of saturation over time \cite{Kiyama2009ExaminationSubstrates}, and over one order of magnitude higher selectivity than the other choice of etchant \cite{Gottmann2017SelectiveSpeed,Raciukaitis2021ChemicalLaser}. However, there will still be some undesired etching of the unmodified material, which we call over-etching (\cref{fig:sle_caveat_overetching}). To mitigate this issue in our fabrication, we extrude the design workpiece with the amount of expected over-etching \cite{Wang2019MultilayeredProcessing}.

\subsection{Depth and resolution trade-off}
In SLE, the lateral resolution of the material modification scales with the numerical aperture (NA) of the objective lens. However, an increased NA results in the reduction of working distance and thus limits the height of the workpiece. The trade-off between the depth and resolution is visualized in \cref{fig:sle_caveat_depth_resolution}. Other considerations include the loss of laser pulse power in the material, which further limits the modifiable substrate depth. Laser power adjustment is needed to achieve constant modified material volume throughout the depth. Laser power correction can also partially increase the depth and resolution up to the maximum achievable power for the laser.

\newpage
\bibliography{references}

\end{document}